\documentclass[acmtog,nonacm]{acmart} %acmtog]{acmart}%,review,anonymous]{acmart}

\usepackage{bm}
\usepackage{bbm}
\usepackage[linesnumbered,ruled,vlined]{algorithm2e}
\usepackage{titlesec}
\usepackage{subcaption}
\usepackage{wrapfig}
\usepackage{svg}
\usepackage{graphicx}
\usepackage{enumitem}
\usepackage{float}
\usepackage{stfloats}

\titleformat{\paragraph}
{\normalfont\normalsize\bfseries}{\theparagraph}{1em}{}
\titlespacing*{\paragraph}
{0pt}{3.25ex plus 1ex minus .2ex}{1.5ex plus .2ex}

\usepackage{xcolor}

\SetCommentSty{mycommfont}

\SetKwRepeat{Do}{do}{while}
\AtBeginDocument{%
  \providecommand\BibTeX{{%
    \normalfont B\kern-0.5em{\scshape i\kern-0.25em b}\kern-0.8em\TeX}}}

\definecolor{peach}{rgb}{ 0.943, 0.188, 0.526}
\definecolor{plum}{rgb}{ 0.858, 0.188, 0.478}
\definecolor{muted_navy_blue}{RGB}{63, 75, 166}
\definecolor{muted_sky_blue}{RGB}{134,166,213}
\definecolor{federal_blue}{RGB}{0,96,240}
\definecolor{regulation_red}{RGB}{226, 20, 79}
\definecolor{federal_gold}{RGB}{240, 212, 14}

\def\figref#1{Fig.~\ref{#1}}

\def\I{\mathbb{I}}
\def\II{\mathbb{II}}
\def\coeff{s}
\def\bcoeff{b}

\DeclareMathOperator*{\diag}{diag}

\begin{document}

%%
%% The "title" command has an optional parameter,
%% allowing the author to define a "short title" to be used in page headers.
\title{Estimating Cloth Elasticity Parameters From Homogenized Yarn-Level Models}
%%\title{Homogenized position based yarn-level cloth}

%%
%% The "author" command and its associated commands are used to define
%% the authors and their affiliations.
%% Of note is the shared affiliation of the first two authors, and the
%% "authornote" and "authornotemark" commands
%% used to denote shared contribution to the research.
\author{Joy Xiaoji Zhang}
\affiliation{%
  \institution{Cornell University, Meta Reality Labs}
  \country{USA}
}

\author{Gene Wei-Chin Lin}
\affiliation{%
  \institution{Meta Reality Labs}
  \country{Canada}
}

\author{Lukas Bode}
\affiliation{%
  \institution{Meta Reality Labs}
  \country{Switzerland}
}

\author{Hsiao-yu Chen}
\affiliation{%
  \institution{Meta Reality Labs}
  \country{USA}
}

\author{Tuur Stuyck}
\affiliation{%
  \institution{Meta Reality Labs}
  \country{USA}
}

\author{Egor Larionov}
\affiliation{%
  \institution{Meta Reality Labs}
  \country{USA}
}

%%
%% By default, the full list of authors will be used in the page
%% headers. Often, this list is too long, and will overlap
%% other information printed in the page headers. This command allows
%% the author to define a more concise list
%% of authors' names for this purpose.
\renewcommand{\shortauthors}{Joy Xiaoji Zhang, et al.}

% \twocolumn[{%
% \renewcommand\twocolumn[1][]{#1}%
% \maketitle
% \begin{center}
\begin{teaserfigure}
\centering
  \includegraphics[width=0.95\textwidth]{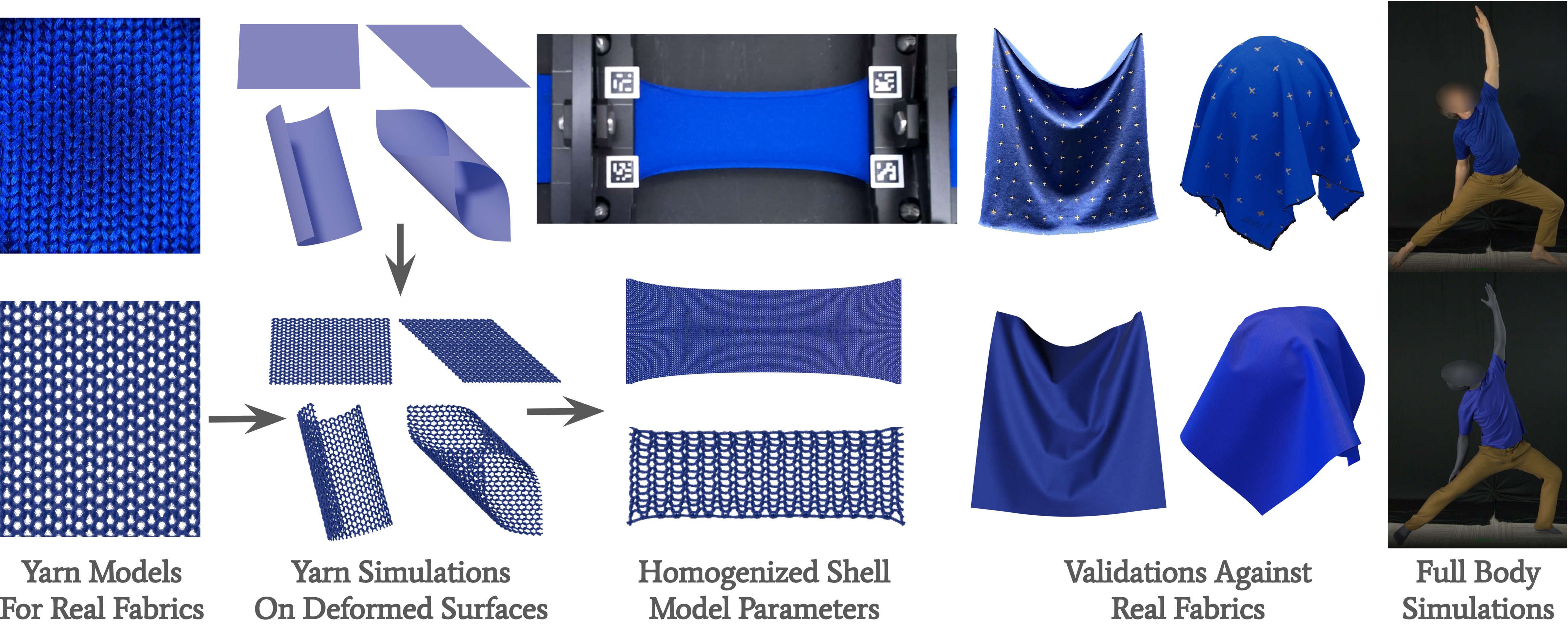}
  \caption{Starting from real fabrics, we construct the matching yarn structures and derive the yarn model parameters. Subsequently, we sample a range of in-plane and out-of-plane deformations, run periodic yarn-level simulations on each deformed surface and record the yarn responses, which we use for optimizing the membrane and bending elasticity parameters of a shell model. We validate our shell parameters by comparing against full yarn-level simulations as well as real fabric captures and measurements. Finally, we simulate full-body garments and compare to reference footage. }
  \label{fig:teaser}
  \end{teaserfigure}
% \end{center}%
% }]
%%
%% The abstract is a short summary of the work to be presented in the
%% article.
\begin{abstract}
    % Virtual clothing is a crucial part of personal expression and character customization in virtual environments.
    % It is important that these garments behave predictably as they do in the real world.
    % This enables captivating visual effects, effective textile design and alluring virtual try-on applications. 
    % Garments are typically modeled and simulated as thin shells, which require expert knowledge and manual parameter tuning.
    % Most real garments have a complex knit or weave structure, which results in various nonlinear and anisotropic behaviors.
    % This makes it difficult to establish a useful shell model that exhibits the same behaviors but is also fast to compute.
    % Yarn-level simulations can reproduce such effects, however, they are more complex and considerably more computationally expensive.
    Virtual garment simulation has become increasingly important with applications in garment design and virtual try-on. However, reproducing garments faithfully remains a cumbersome process. We propose an end-to-end method for estimating parameters of shell material models corresponding to real fabrics with minimal priors. Our method determines yarn model properties from information directly obtained from real fabrics, unlike methods that require expensive specialized capture systems. We use an extended homogenization method to match yarn-level and shell-level hyperelastic energies with respect to a range of surface deformations represented by the first and second fundamental forms, including bending along the diagonal to warp and weft directions. We optimize the parameters of a shell deformation model involving uncoupled bending and membrane energies. This allows the simulated model to exhibit nonlinearity and anisotropy seen in real cloth. Finally, we validate our results with quantitative and visual comparisons against real world fabrics through stretch tests and drape experiments. Our homogenized shell models not only capture the characteristics of underlying yarn patterns, but also exhibit distinct behaviors for different yarn materials. %show that in addition to the underlying yarn structures, physical properties such as density of yarns and material type also affect the estimation of shell model parameters.
\end{abstract}

%%
%% The code below is generated by the tool at http://dl.acm.org/ccs.cfm.
%% Please copy and paste the code instead of the example below.
%%
% \begin{CCSXML}
% <ccs2012>
%    <concept>
%        <concept_id>10010147.10010371.10010352.10010379</concept_id>
%        <concept_desc>Computing methodologies~Physical simulation</concept_desc>
%        <concept_significance>500</concept_significance>
%        </concept>
%    <concept>
%        <concept_id>10010405.10010432.10010441</concept_id>
%        <concept_desc>Applied computing~Physics</concept_desc>
%        <concept_significance>300</concept_significance>
%        </concept>
%  </ccs2012>
% \end{CCSXML}

% \ccsdesc[500]{Computing methodologies~Physical simulation}
% \ccsdesc[300]{Applied computing~Physics}

%%
%% Keywords. The author(s) should pick words that accurately describe
%% the work being presented. Separate the keywords with commas.
% \keywords{yarn-level cloth simulation, homogenization}

% \received{20 February 2007}
% \received[revised]{12 March 2009}
% \received[accepted]{5 June 2009}

%%
%% This command processes the author and affiliation and title
%% information and builds the first part of the formatted document.
\maketitle

\section{Introduction}
%Questions we should answer
%\begin{itemize}
%\item What problem are we solving? (define scope)
%\item why are we solving it? Why is it important?
%\item how is it different than previous work?
%\end{itemize}

%\todo{Do a pass over the introduction to well position our work with respect to related work. Set expectations early : Our goal is not to produce the most accurate shell level material models but instead, we get pretty good estimates using very little information. Estimates can be refined using other methods if needed. Add use cases of providing good initial estimate for other material optimization pipelines that do use more expensive and harder to obtain measured data.}

The physical realism of simulated cloth plays a critical role in digital garment design. Cloth simulation, once primarily used in video games and animations, has now become a crucial aspect of this creative process. Increasingly, faithful reproduction of clothing behaviors has become a focus point in garment design software \cite{CLO3D, Vstitch, style3D} where the designed textiles are simulated to determine the fit on a virtual body. %The fashion industry is known for its fast-paced nature, with new trends and styles emerging quickly, which motivates manufacturers to produce garments rapidly. However, garment design and creation can have long lead times involving several complex stages. 
%Virtual fitting applications allow for more rapid design iterations and have the potential to greatly reduce the time and effort spent when manufacturing garments.
The simulator needs to accurately predict the behaviors of different types of fabrics composed of numerous knitted or woven yarn structures consisting of various materials. While \cite{wu2020weavecraft} demonstrate simulation-assisted design of complex 3D patterns such as a woven shoe, modeling and simulating faithful full body garments that account for the underlying fabric compositions remains challenging. 

Textile design software commonly model fabrics as linear or hyperelastic materials \cite{CLO3D}. While it is possible for shell simulations to be computed in real-time on the GPU, they generally require tedious manual processes and expert knowledge in tuning the parameters to reproduce satisfactory dynamics of a desired material type. In addition, the limited degrees of freedom in shell models makes it exceedingly hard for shell simulations to exhibit the rich behaviors that we observe in real fabrics such as edge curling in stockinette patterns and anisotropic bending in weave patterns.

Several approaches have been explored to produce more faithful cloth behaviors. One solution is to model cloth as a collection of interacting yarns \cite{kaldor2008simulating,leaf2018interactive}. This approach is quite accurate in reproducing the complex cloth behaviors but at an excessively high computational cost. Another line of research focuses on improving the shell model and gaining better automatic estimations for the model parameters. A handful of works \cite{larionov2022estimating, miguel2012data, clyde2017modeling, wang2011data, feng2022learning} estimate the cloth elasticity parameters from a range of data gathered from real fabrics including physical experiments of mechanical properties and high resolution images. Although these methods are capable of producing high-quality results, they require potentially expensive specialized equipment. On the other hand, \citet{sperl2020hylc} learns shell material models from yarn-level deformation responses using numerical homogenization, which is closely related to our work. %Our work builds on a similar homogenization technique, but instead optimizes the parameters of predefined thin shell models, which require less investigation of the collected yarn responses with regards to interactions between different parameters. Furthermore, we account for the effect of yarn material properties in addition to yarn patterns, and evaluate our parameter estimates by comparing simulated shells against physical tests of real fabrics. \todo{Move the last sentence to the next paragraph}%\todo{Highlight benefits of doing it this way.}

We present a method for estimating the shell-level cloth material parameters required to simulate real fabrics given only basic information about the fabric.
Using physics-based simulations, we provide parameter estimations that reflect the behaviors of various fabric compositions.
% Through the use of physics-based simulation and numerical homogenization, we are capable of providing parameter estimations that exhibit the representative behaviors of fabrics composed of various yarn structures and materials.
%Our method achieves this without complex priors such as measurements using specialized equipment and inverse pipelines involving differentiable simulation. %\joy{is this enough for addressing the core contribution?} %including material type, yarn structure and yarn density, typically found in garment labels and simple measurements
Given a real fabric, we derive the yarn model parameters from simple measurements as well as freely available experimental data from textile research, and use numerical homogenization \cite{sperl2020hylc} for collecting yarn-level responses to a range of shell deformations including anisotropic bending, while accounting for the nonlinearity in yarn stretching. %Our method takes these into account to obtain representative yarn-level simulation parameters. %We propose a mapping from fabric properties to yarn simulation parameters.
%We then define a range of in-plane and out-of-plane surface deformations. For each sampled deformation, we periodically tile a yarn pattern on a corresponding deformed surface. We then simulate the yarns to a relaxed state and record the yarn material responses. Given these material responses under different deformations, we optimize the material parameters of shell-level models, enabling efficient simulation of the fabrics.\todo{Simplify this and only explain the part that's different from \citet{sperl2020hylc}}
Our periodic yarn relaxation result (see \figref{fig:pipeline}) can further drive an efficient surface-based cloth appearance model~\cite{zhu2023realistic}, which drastically
enhances rendered results.
% Beyond being a stepping stone towards efficient shell simulation, the relaxed yarns can be used to drive an efficient surface-based cloth appearance model~\cite{zhu2023realistic}.

% \todo{ In addition to our material estimation pipeline, we propose a nonlinear Saint-Venant Kirchhoff (StVK) thin shell model that reproduces the nonlinear stretching response more faithfully compared to the orthotropic linear StVK model}.
In contrast to previous work that learns a different material model for every yarn pattern~\cite{sperl2020hylc}, we optimize the thin shell model parameters directly and demonstrate that the parameters are not only determined by the yarn patterns, but also the physical properties such as material type and yarn thickness. Furthermore, we estimate the off-diagonal terms in the bending matrix to capture anisotropic bending effects, thus constructing the entire stiffness matrix through homogenization and eliminating the laborious process of tuning parameters by hand. Finally, we validate our homogenized shell parameters against machine measurements of real fabric swatches, and discover that the simulated fabrics are not only capable of visually establishing the representative behaviors of each material, but also match the real fabrics quantitatively despite the absence of an inverse process in our pipeline.

Our approach generalizes to shell models that are based on the first two fundamental forms (i.e. stretching and bending strains). In addition, we estimate bending parameters along all three out-of-plane directions, providing reliable initial estimates for more complex parameter estimation methods, which can improve over our estimates using specialized machines. % where the bending measurements cannot be easily obtained. %\joy{just added this line} %In addition to modeling bending in the warp or weft directions separately, we model anisotropic bending by estimating off-diagonal terms in the bending stiffness matrix. 

%[\joy{Hybrid approach} - Compute/Approximate yarn simulation parameters directly from physical properties, then homogenize (simulate yarns and gather deformation-response data)]

%[Benefit: Can obtain the shell parameters for arbitrary clothing, as long as we know the type of material, the yarn pattern, and the density in mass per unit area. We don't even need a FAB machine!]

%[We need to know one of the following sets of information:
%[1. Material type, pattern, (Young's modulus and density), Poisson's ratio]
%[2. Material type, pattern, (stiffness and denier), Poisson's ratio]

%[Denier -> radius]
%[Density -> radius]
%[Stiffness -> Young's modulus]

%List of contributions:
%\begin{itemize}
%\item Homogenization pipeline for common material models like StVK.
%\item Anisotropic bending homogenization.
%\item Improved material models. (how are they better?)
%\item (bonus) Damping estimation?
%\end{itemize}

%We present a method for producing yarn-level clothing behaviors through simulating sheet models. Similar to \citet{sperl2020hylc}, we first gather yarn responses to surface-level deformations for a range of knitted and woven patterns. However, instead of learning an approximate model from the yarn data via spline fitting, we directly fit the stiffness parameters of linear and nonlinear Saint-Venant Kirchhoff (StVK) models.

To summarize, we offer the following technical contributions, which allow us to model shell-level cloth starting from only a basic fabric description:
\begin{itemize}
\item An end-to-end method to estimate shell simulation parameters corresponding to real fabrics that is compatible with any shell material model based on the first and second fundamental forms. %\todo{Agnostic to material model? i.e. also applicable for non stvk models?} %via numerical homogenization from yarn models
\item A novel yarn parameter estimation method from simple measurements of real fabrics, with a nonlinear stretching model incorporating experimental results of textile research. %direct mapping from measured fabric and/or individual yarn properties to \egor{yarn-level <- shell-level?} \joy{Yarn level} simulation properties
\item An approach for homogenizing anisotropic bending including the warp-weft coupling term.
% \item An extended StVK thin shell model to capture nonlinear deformation responses that is compatible with compliant constraint dynamics.%rich effects such as in-plane compression of garments
\end{itemize}

\section{Related Work}

\textbf{Shell-Level Cloth Simulation.} Starting with the pioneering work of \citet{baraff98}, thin shell cloth simulations have seen widespread success over the last decades with applications in animation and special effects \cite{stuyck2022cloth}. Since then, many improvements have been made relating to robustness, accuracy and efficiency. \citet{muller2007position} proposed a Position Based Method (PBD) designed for efficiency where constraints are solved in parallel by updating positions directly. To overcome the limitations of PBD, eXtended Position Based Dynamics (XPBD) \cite{miles2016xpbd} was proposed to eliminate iteration count dependency of simulation results. \citet{bouaziz14pd} employed an implicit integrator which bridges the gap between continuum mechanics and PBD. In contrast to prior linear element methods, \citet{ni2023simulating} simulated thin shells using the bicubic Hermite element method.

\textbf{Yarn-Level Cloth Simulation.} Shell-level cloth has seen many advances in recent years which significantly improve realism, yet it remains difficult to reproduce all intricate behaviors of fabrics. To improve on these models, \citet{kaldor2008simulating} proposed original work to model knits at the yarn-level.
Generating yarn-level geometries can be cumbersome. To alleviate this, \citet{yuksel2012stitch} created full garment yarn-level geometries based on simple triangle models.
% It is cumbersome to generate these yarn-level geometries, to alleviate this, \citet{yuksel2012stitch} generated full garment yarn-level geometries based on easily modeled triangle models.
Albeit more computationally expensive, they were capable of reproducing unique effects such as edge curling under tension. \citet{cirio2014yarn} extended this work with a focus on computational efficiency by modeling interlaced yarns based on yarn crossings and yarn sliding, with implicit contacts and persistent contacts in follow-up work \cite{cirio2015efficient, cirio2016yarn}. A different way to improve efficiency is to model the yarns periodically \cite{leaf2018interactive}, enabling interactive pattern design. In order to obtain yarn-level like behavior at moderate computational cost, \citet{casafranca2020mixing} combined both shell-level and yarn-based simulation models in a single framework.

Numerous yarn representations have been proposed. \citet{remion1999dynamic} developed equations for modeling knitted patterns with spline curves, and uses springs between control points for length preservation. \citet{jiang2005geometric} modeled the relaxed configuration of woven patterns using spline curves. There has also been a series of work on modeling and simulating thin elastic rods based on Kirchhoff rods \cite{bergou2008discrete, bertails2006super, spillmann2007corde} that uses discrete poly-line representations. Our yarn-level simulation employs the Cosserat rod model \cite{spillmann2007corde} and the periodic boundary conditions \cite{leaf2018interactive}.
% \todo{Include remaining references}
% \begin{itemize}
%     \item Springs
%     \item Splines
%     \item Cosserat rods
%     %\item Simulating knitted cloth at the yarn level \cite{kaldor2008simulating}
%     %\item Yarn-level simulation of woven cloth \cite{cirio2014yarn}
%     %\item Efficient simulation of knitted cloth using persistent contacts \cite{cirio2015efficient}
%     %\item Yarn-level cloth simulation with sliding persistent contacts \cite{cirio2016yarn}
%     %\item Mixing yarns and triangles in cloth simulation \cite{casafranca2020mixing}
%     %\item Stitch meshes for modeling knitted clothing with yarn-level detail \cite{yuksel2012stitch}
%     %\item Interactive design of periodic yarn-level cloth patterns \cite{leaf2018interactive}
%     \item Mechanical characterization of structured shell materials \cite{schumacher2018mechanical}
% \end{itemize}

\textbf{Multiscale Modeling and Homogenization of Cloth.} In computer graphics, representing cloth using a mixture of scales has been extensively studied to leverage the efficiency of simulating shell models while preserving the detailed effects produced by yarn models. \citet{martin2010unified} defined a unified, high-order integration for elastic responses shared among elastic rods, shells and solids. \citet{casafranca2020mixing} proposed a kinematic transition between triangle and yarn representations. \citet{fei2018multi} produced anisotropic effect in cloth-fluid coupling by modeling the fiber-level porous structures. Our work is most closely related to \citet{schumacher2018mechanical} and \citet{sperl2020hylc} that derived shell models using elastic responses of underlying thin structures subjecting to shell deformations. \citet{sperl2022} presented a technique for modeling the yarn-level mechanics of cloth, based on real fabric responses to mechanical forces. \citet{sperl2021} animated yarn-level cloth geometry on top of a deformed mesh while accounting for the underlying mechanics.

% \todo{}
% \begin{itemize}
%     \item HYLC \cite{sperl2020hylc} and previous work.
%     \item Applications of homogenized cloth for yarn property estimation and cloth yarn rendering \cite{sperl2021, sperl2022}
% \end{itemize}
%\todo{Are we missing references to \cite{sperl2021, sperl2022} ?}

\textbf{Material Parameter Estimation.} Several works focus on estimating material models and parameters to model clothing based on captured data. Early work optimized material parameters from videos of captured cloth \cite{bhat}. Follow up work captured cloth deformations under known loads to estimate the material models \cite{miguel2012data, clyde2017modeling, wang2011data}. \citet{larionov2022estimating} proposed the use of an FFT-based loss to robustly handle bifurcations in cloth buckling. Differentiable simulation has successfully been applied to material estimation from synthetic \cite{stuyck2022cloth} and real data \cite{li2023diffavatar}. \citet{wang2020learning} learned a material model directly instead of fitting parameters to existing material models.

\section{Background} \label{sec:background}
%\begin{itemize}
%    \item Brief XPBD summary
%    \item cosserat rods and cloth simulation overview
%    \item homogenization in the context of XPBD
%\end{itemize}

%The main goal of our system is to estimate the thin shell model parameters to simulate arbitrary real world fabrics with minimal physical measurements. 

%Previous work have focused on estimating parameters of a target thin shell model through a set of experimental measurements of mechanical properties as well as high-resolution image captures \cite{sperl2022,egor}. While such approaches

%We propose a nonlinear StVK model in addition to the widely used orthotropic StVK thin shell model, and discuss how the nonlinear model captures a wider array of clothing behaviors. Estimating yarns from machine data: Hard to 

We briefly review a few key concepts on which our parameter estimation pipeline is based. We introduce the Cosserat rod model for representing yarns, the orthotropic StVK material for modeling thin shells, as well as numerical homogenization detailed by \citet{sperl2020hylc}. We expand on XPBD, a constraint-based simulation framework, that we employ for both the yarn-level and shell-level simulations.

Although we use constraint based dynamics, our approach is compatible with any solver that finds the static equilibrium of yarns.
% It is important to note that while we devise our model as a set of constraints to conform with the XPBD formulation, our approach can be extended to any type of solver that finds the static equilibrium of yarns. 
Our technical contribution lies in deriving strain-dependent yarn model parameters from freely available measurements of real yarns and fitting shear bending terms to homogenized yarn-level cloth, which was omitted in previous work.

\subsection{Compliant Constraint Formulations of Models}
%\el{There is an inconsistency in how we write the energy in yarn and shell eqs, e.g. eq (3) vs eq (7), where the latter assumes $A$ is part of $\alpha$, where as in (3) $l_0$ is external.}
%\gene{thanks for pointing out!}
For both yarn and shell simulations, we follow \citet{miles2016xpbd} and minimize an energy potential of the form
\begin{equation}
\label{eqn:xpbd}
U(\bm{x}) = \frac 12 \bm{C}(\bm{x})^{\top} \boldsymbol{\alpha}^{-1} \bm{C}(\bm{x})
\end{equation}
where \(\boldsymbol{\alpha}\) is a compliant matrix that represents the inverse stiffness and \(\bm{C}\) are bilateral constraint functions describing the kinematics of the dynamic system \cite{bender2015position}. We describe shell-level constraints with \(\overline{\bm{C}}\) to distinguish from yarn-level constraints \(\bm{C}\). %\joy{is this enough?} \gene{It's obvious for us but should we describe what $\bm{C}(\bm{x})$ is?}

\subsection{Modeling Yarns With Cosserat Rods}

%\todo{Explain all symbols}

A Cosserat rod is described by a centerline curve and a set of local orthonormal frames \((\bm{d}_1,\bm{d}_2,\bm{d}_3)\) attached to each point on the curve, such that \(\bm{d}_3\) aligns with the centerline direction and \((\bm{d}_1,\bm{d}_2)\) span the cross-sectional direction of the rod.

We represent the yarn centerline as a set of points \(\{\bm{p}_0,\dots,\bm{p}_N\}\) and edges \(\{\bm{e}_0,\dots,\bm{e}_{N-1}\}\) where \(\bm{e}_i=\bm{p}_{i+1}-\bm{p}_{i}\). Each edge has an associating local frame written as a quaternion $\bm{q}$ that consists of a real part \(q_0=\Re(\bm{q})\) and an imaginary part \(\bm{q}_{im}=\Im(\bm{q})\), representing the angle of rotation \(\theta=\cos^{-1}(q_0)\) around an axis \(\frac{\Im(\bm{q})}{\sin\theta}\). In our yarn model, \(\Im_0, \Im_1\) represent bending and \(\Im_2\) corresponds to twisting. Contrary to the Discrete Elastic Rod formulation \cite{bergou2008discrete} where the bending and twisting are calculated from \((\phi, \theta)\) corresponding to the turning angle between adjacent edges and the rotation angle of local frames, the quaternion is a unified representation of both the bending and twisting, i.e. \(\bm{q}^T=[q_0,\bm{q}_{im}^T]^T\) such that \(\bm{R}\left(\bm{q}\right) = [\bm{d}_1,\bm{d}_2,\bm{d}_3]\), and requires additional geometric constraints to align the local frames with the centerline. The Rodriguez formula \(\bm{R}\left(\bm{q}\right)\) computes the rotation from world frame to local frame.
\begin{equation}
\label{eqn:rodriguez}
\begin{aligned} 
\bm{R}\left(\bm{q}\right) &= 2\bm{q}\bm{q}^T + \left(q_{0}^2-\bm{q}_{im}^T\bm{q}_{im}\right)\bm{1} + 2q_{0}\left[\bm{q}_{im}\right]^\times \\ 
\end{aligned}
\end{equation}

Our yarn model follows the formulation by \citet{kugelstadt2016position} where we model stretching-shearing and bending-twisting energies. The total energy for a yarn with \(N+1\) points is
%Our yarn model follows the formulation by \citet{kugelstadt2016position}. We model the stretching/shearing energies exhibited by all adjacent points and the bending/twisting energies exhibited by all adjacent edges. The total energy for a yarn with \(N+1\) points is %, alongside additional energies induced in the homogenization procedure, which we elaborate in \todo{\ref{sec:homogenized_yarns}}.
%Given a discretized yarn with sampled points \(\{\bm{p}_0,\dots,\bm{p}_N\}\) and quaternions \(\{\bm{q}_0,\dots,\bm{q}_{N-1}\}\), we write the energy as
\begin{equation}
\label{eqn:yarn_energy}
\begin{aligned}
U_{\text{yarn}}&=U_{\text{stretch\_shear}}+U_{\text{bend\_twist}}\\
&=\sum_{i=0}^{N-1} \frac{1}{2}\bm{C}_s^T\bm{\alpha}_s^{-1}\bm{C}_s + \sum_{i=0}^{N-2}\frac{1}{2}\bm{C}_b^T\bm{\alpha}_b^{-1}\bm{C}_b\ \text{where}\\
\boldsymbol{\alpha}_s^{-1} &= \begin{bmatrix}EA & & \\ & EA\\ & & EA\end{bmatrix}l_0,\ \ 
 \boldsymbol{\alpha}_b^{-1}= \begin{bmatrix} E I_1 & & \\ & E I_2 & \\ & & G I_3\end{bmatrix}l_0.
\end{aligned}
\end{equation}
% \gene{There's no definition of $\alpha_s^{-1}$ and $\alpha_b^{-1}$ until eq (12). Do we need to move eq (12) and its description to here and briefly mention eq (3) in section 4.1?}
% \joy{Will do! Thanks Gene!}
% For an arbitrary yarn segment \((\bm{p}_i,\bm{p}_{i+1})\) with rest length \(l_0\),
% \begin{equation}
% \label{eqn:rod_stiffness}
% \begin{aligned} 
% \end{aligned}
% \end{equation}
Here, \(E\) is the Young's modulus, \(G\) is the shear modulus where \(E = 2G(1+\nu)\), \(\nu\) is the Poisson's ratio, and \(I_i\) is the second area moment of inertia. For a cylindrical yarn with circular cross-section, \(A = \pi r^2\), \(I_1 = I_2 = \frac{1}{4} \pi r^2\) and \(I_3 = \frac{1}{2}\pi r^2\).

\noindent \textbf{Stretching-Shearing} %The constraint
%\vspace{-0.4cm}
\begin{equation}
\begin{aligned} 
\label{eqn:stretch_shear}
\bm{C}_s\left(\bm{p}_{i}, \bm{p}_{i+1}, \bm{q}_i\right) &= \frac{1}{l_0} \left(\bm{p}_{i+1} - \bm{p}_{i}\right) - \bm{d}_3(\bm{q}_i)\\
&= \frac{1}{l_0} \left(\bm{p}_{i+1} - \bm{p}_{i}\right) - \bm{R}(\bm{q}_i)\hat{\bm{e}}_3,
\end{aligned}
\end{equation}
where \(l_0\) is the rest length of edge $\bm{e}_i$ and \(\hat{\bm{e}}_3\) is a unit vector in the yarn center-line direction in world frame.\\
%\bm{R}\left(\bm{q}\right)\bm{e}_{3}, \\ 
%where \(\bm{R}\left(q\right)\) is the Rodriguez formula which relates the rotation matrix \(\bm{R}\) and the quaternion \(q\)
\noindent \textbf{Bending-Twisting} %The constraint
\vspace{-0.3cm}
\begin{equation}
\begin{aligned}
\label{eqn:bend_twist}
\bm{C}_b\left(\bm{q}_i, \bm{q}_{i+1}\right) &= \bm{\Psi} - \bm{\Psi}^{0} 
= \frac{2}{l_0}\frac{\Im\left(\bar{\bm{q}}_i\bm{q}_{i+1}\right)}{\Re\left(\bar{\bm{q}}_i\bm{q}_{i+1}\right)} - \frac{2}{l_0}\frac{\Im\left(\bar{\bm{q}}^{0}_i\bm{q}^{0}_{i+1}\right)}{\Re\left(\bar{\bm{q}}^{0}_i\bm{q}_{i+1}^{0}\right)},
\end{aligned}
\end{equation}
%\vspace{-0.5cm}
where the superscript 0 denotes the rest configuration and $\bm{\Psi}$ is the modified Darboux vector.
% equals %\todo{Both the Darboux vector and the StVK energy use the \(\bm{\Psi}\) symbol! Need a different symbol here?}
% \begin{equation}
% \begin{aligned}
% \label{eqn:darboux}
% \bm{\Psi} &= \frac{2}{l_0}\Im\left(\bar{\bm{q}}_{i}\bm{q}_{i+1}\right)/\Re\left(\bar{\bm{q}}_{i}\bm{q}_{i+1}\right)
% \end{aligned}
% \end{equation}

\subsection{Modeling Shells with Orthotropic StVK}
\label{sec:shell_stvk}
To simulate hyperelastic shells, we require membrane (in-plane) and bending (out-of-plane) energies.
We focus on 2D knitted and woven patterns where the yarn patterns are aligned with two orthogonal directions, which we henceforth refer to as \emph{weft} and \emph{warp} indicated by subscripts ${}_t$ and ${}_p$ respectively\footnote{Note that the terms warp and weft apply only to woven fabrics in the textile industry.}. We compute the strain following the discrete Kirchhoff-Love shell model~\cite{weischedel12, chen2018thinShell,Wen2023Shell}. The in-plane stretching strain can be represented as %constraint
%$\tilde\epsilon = (\epsilon_{tt}, \epsilon_{pp}, 2\epsilon_{tp})^{\top}$, where
\begin{equation}
\label{eqn:stretch_strain}
\overline{\boldsymbol{\epsilon}}_s = \begin{bmatrix} \epsilon_{tt} & \epsilon_{tp} \\ \epsilon_{tp} & \epsilon_{pp} \end{bmatrix} \propto \overline\I_{\text{rest}}^{-1}(\overline\I - \overline\I_{\text{rest}})
\end{equation}
where $\overline\I$ and $\overline\I_{\text{rest}}$ are the first fundamental forms on the deformed and rest surfaces respectively. The elements $\epsilon_{tt}$, $\epsilon_{pp}$ and $\epsilon_{tp}$ measure weft, warp and shear deformations respectively.
Let us define $\tilde{\I} = \overline\I_{\text{rest}}^{-1}(\overline\I - \overline\I_{\text{rest}})$, then the constraint and stiffness matrix are%stiffness matrix is%$\bm{\alpha}^{-1}_\stvk$ is
\begin{equation}
\label{eqn:shell_stiffness}
\begin{aligned}
%\overline{\bm{C}}_s&=\sqrt{A}\overline{\boldsymbol{\epsilon}}_s\\
\overline{\bm{C}}_s&=\sqrt{A}(\tilde{\I}_{tt}, \tilde{\I}_{pp}, \tilde{\I}_{tp}) \\
\overline{\bm{\alpha}}^{-1}_s &= h \begin{bmatrix} \coeff_{00} & \coeff_{01} & \\ \coeff_{01} & \coeff_{11} & \\ & & \coeff_{22} \end{bmatrix}
\end{aligned}
\end{equation}
where $A$ is the area of a triangle, and $h$ is the thickness of the shell. The weft and warp stiffness of the fabric are controlled by $\coeff_{00}$ and $\coeff_{11}$ respectively. $\coeff_{01}$ controls the coupling between deformations in warp and weft directions, and $\coeff_{22}$ is the shear stiffness.
We can express the stiffness values in terms of the more intuitive Young's moduli \(E_p, E_t\), Poisson's ratios \(\nu_{pt}, \nu_{tp}\) and shear modulus \(\mu\):
\begin{equation}
\label{eqn:shell_stiffness_moduli}
\begin{aligned}
\begin{bmatrix} \coeff_{00} & \coeff_{01} & \\ \coeff_{01} & \coeff_{11} & \\ & & \coeff_{22} \end{bmatrix} = 
\frac{1}{1-\nu_{tp}\nu_{pt}}\begin{bmatrix} E_t & \nu_{pt}E_t & \\ \nu_{tp}E_p & E_p & \\ & & \mu(1-\nu_{tp}\nu_{pt}) \end{bmatrix}.
\end{aligned}
\end{equation}
%This matrix is symmetric because \(\nu_{vu}E_u = \nu_{uv}E_v\). \(\nu_{xy}\) is the Poisson's ratio that corresponds to a contraction in direction \(y\) when an extension is applied in direction \(x\).

The bending strain is proportional to the change in the curvature of the surface $\tilde{\II} = \overline\I_{\text{rest}}^{-1}(\overline\II - \overline\II_{\text{rest}})$, where $\overline{\II}$ and $\overline{\II}_{\text{rest}}$ are the second fundamental forms. Using the same formulation, we define the bending constraints as
\begin{equation}
\label{eqn:bend_strain}
\begin{aligned}
\overline{\bm{C}}_b =\sqrt{A}(\tilde{\II}_{tt}, \tilde{\II}_{pp}, \tilde{\II}_{tp})
\end{aligned}
\end{equation}
All our experiments assume a flat rest shape with $\overline{\II}_{\text{rest}}=\bm{0}$. The stiffness matrix is \(\overline{\bm{\alpha}}^{-1}_b = h^3 \diag(\bcoeff_{00}, \bcoeff_{11}, \bcoeff_{22})\).
%, and the constraint \(\overline{\bm{C}}_b=\sqrt{A}\overline{\boldsymbol{\epsilon}}_b\). 
%We use distinct stiffness matrices for membrane and bending energies and formulate constraints for XPBD. This bending model can represent a large variety of aesthetics as demonstrated in \figref{fig:bend_demo}.
\begin{figure}[h]
    \vspace{-0.3cm}
    \centering
    \includegraphics[width=0.8\linewidth]{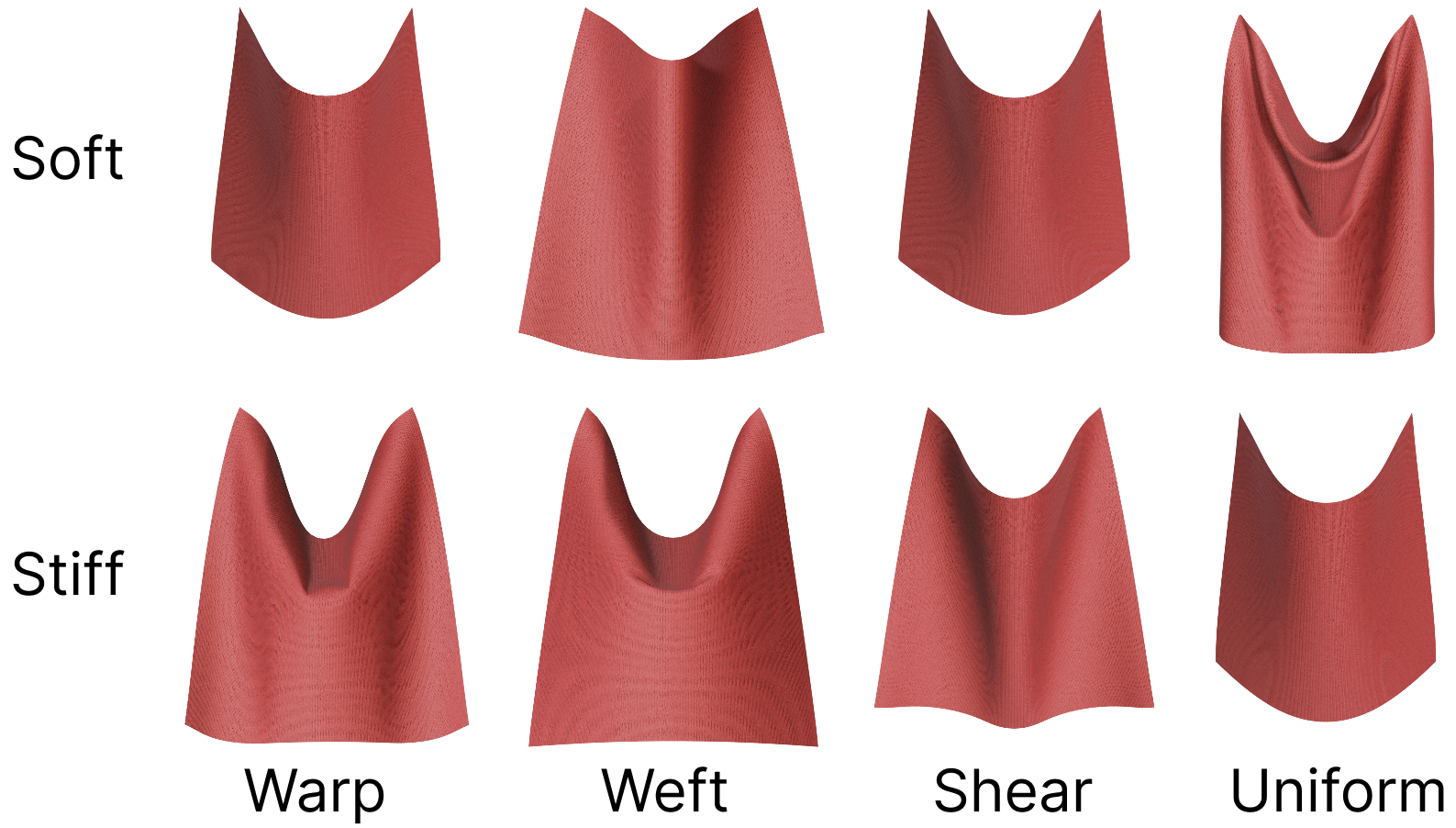}
    \caption{A square swatch (oriented vertically along the warp direction) is draped using various bending parameters. Each column indicates which parameter is modified with respect to the others. The right column makes all parameters either soft (top row) or stiff (bottom row). \label{fig:bend_demo}}
    \vspace{-0.4cm}
\end{figure}

\subsection{Homogenization}
%\todo{Explain RVE}
Homogenization aims at obtaining macroscopic equations for systems with a fine microscopic structure. Following \citet{sperl2020hylc}, we describe the macroscopic thin shell using reference coordinates \(\overline{\bm{X}}\), deformed coordinates \(\overline{\bm{x}}\), and deformation gradient \(\overline{\bm{F}}=\frac{\overline{\bm{x}}}{\overline{\bm{X}}}\). Correspondingly, the microscopic yarn-level structure has the descriptors \(\bm{X}\), \(\bm{x}\) and \(\bm{F}\). %that the macroscopic material exhibits fluctuations on the micro scale that averages over the \textit{representative volume element} of the macro scale.

% The thin shell deformations are defined by the first fundamental form \(\overline{\mathbb{I}}\) for in-plane deformations and the second fundamental form \(\overline{\mathbb{II}}\) for bending deformations.
Given an arbitrary deformation, the deformed thin shell is defined on a mid-surface \(\overline{\boldsymbol{\phi}}\) corresponding to the \textit{representative volume element} (RVE) as $\overline{\bm{x}}=\overline{\boldsymbol{\phi}}(\overline{\bm{X}})+h\overline{\bm{n}}$.
%\begin{equation}
%\label{eqn:shell_rve}
%\overline{\bm{x}}=\overline{\boldsymbol{\phi}}(\overline{\bm{X}})+h\overline{\bm{n}}
%\end{equation}
Here, \(h\) is the thickness or deviation from the mid-surface and \(\bm{n}\) is the surface normal.
The underlying yarn structure undergoes the same deformation, i.e. \(\mathbb{I}=\overline{\mathbb{I}}\), \(\mathbb{II}=\overline{\mathbb{II}}\). The resulting yarn geometry is a subdivision of the macro-scale mid-surface \(\boldsymbol{\phi}\) plus additional fluctuations \(\tilde{\bm{u}}\) induced by the yarn relaxation process expressed as $\bm{x}=\boldsymbol{\phi}(\bm{X})+h\bm{n}+\tilde{\bm{u}}(\bm{X})$.
 %\todo{Illustration. First row: \(Omega\). Second row: Eqn \ref{eqn:shell_rve}. Third row: Eqn \ref{eqn:yarn_rve}}
%\begin{equation}
%\label{eqn:yarn_rve}
%\bm{x}=\boldsymbol{\phi}(\overline{\bm{X}})+h\bm{n}+\tilde{\bm{u}}(\bm{X})
%\end{equation}
Homogenization theory assumes that the macroscale quantities \(\overline{\bm{x}}\) and \(\overline{\bm{F}}\) are averaged from the microscale quantities \(\bm{x}\) and \(\bm{F}\) which span the macro region. In addition, the RVE is treated as constant across the micro region. These two assumptions indicate that the yarn-level fluctuations always average out over the shell mid-surface which is expressed as
\begin{equation}
\label{eqn:rve_constraint}
\begin{aligned}
\int_{\Omega} \tilde{\bm{u}}(\bm{X})d\Omega &= 0, \ \int_{\Omega} \nabla\tilde{\bm{u}}(\bm{X})d\Omega &= 0
\end{aligned}
\end{equation}

\section{Method} \label{sec:method}
\vspace{-0.5cm}
\begin{figure}[H]
    %\hspace{-0.5cm}
    %\vspace{-0.5cm}
    %\centering
    %\includesvg[width=0.5\textwidth]{img/method-4.svg}
    %\includesvg[width=0.5\textwidth]{img/pipeline.svg}
    %\includegraphics[width=0.5\textwidth]{img/pipeline-reduced.pdf}
    \includegraphics[width=0.5\textwidth]{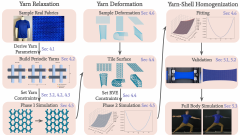}
    %\hspace{-0.5cm}
    %\vspace{-0.45cm}
    %\vspace{0.2cm}
    \caption{Overview of our proposed yarn-shell parameter estimation method.}
    \label{fig:pipeline}
    %\vspace{-0.2cm}
\end{figure}
We now elaborate our end-to-end method for estimating shell simulation parameters corresponding to real fabrics. Our method consists of three stages: (i) initial yarn relaxation, (ii) yarn simulation on deformed surfaces and finally (iii) shell model parameter fitting. Figure~\ref{fig:pipeline} provides a visual overview.
\begin{enumerate}[leftmargin=0em]
  \item[]\textbf{Yarn Initialization and Relaxation.} Given a real garment, we first obtain the material type through information readily found on their labels. We then visually determine the yarn structure through close-up captures and sample a fabric swatch to measure the density. With additional information provided by textile research on the tensile strength of materials, we derive the yarn model parameters. We leverage a database of knitted or woven patterns \cite{leaf2018interactive} to relax the corresponding yarn pattern using periodic boundary conditions (PBC).
  \item[]\textbf{Yarn Simulation on Deformed Surfaces.} We simulate the relaxed yarns to establish the shape of a deformed mid-surface, required for homogenization. Given a sampled deformation at the macro scale, we compute the resulting shape of the shell. Subsequently, we tile yarns on the deformed shell and record the resulting yarn-level energy by running a yarn-level simulation subject to homogenization constraints. This provides a mapping from the deformation space at the macro level to the energy densities at the micro level, which exhibits rich physical responses that are not well represented using the macro level constitutive model.
  \item[]\textbf{Shell Parameter Estimation.} Finally, we fit an orthotropic StVK shell model to the collected yarn-level deformation response data. We validate our homogenized shells for each material against quantitative measurements of real fabrics as well as full yarn simulations (without PBC) on stretching and draping experiments. In addition, we simulate full-sized garments using our homogenized shell parameters and compare with recorded real-world footage.
\end{enumerate}
\subsection{Deriving Yarn Model Parameters from Real Fabrics}
\label{sec:yarn_param_estimation}
Provided with simple measurements of real fabrics, we estimate the stiffness matrices in Eq.~\ref{eqn:yarn_energy} using a nonlinear stretching model inspired by \cite{sperl2022}. %\gene{Do we need to cite the source of \(\alpha_c\)? Alternatively can we remove this constant as it's not in most DER papers, as well as in our implementation?}
% We refer to the hair Cosserat rod model for the XPBD stretch/shear and bend/twist constraints. Our compliance matrices are
% \[
% \begin{aligned}
%     \alpha_{s}^{-1} &= S * l_0\\
%     \alpha_{b}^{-1} &= B * l_0
% \end{aligned}
% \]
% where \(l_0\) is the rest length of the edge corresponding to the constraints.
%In our yarn simulator, the length unit is in millimeter (mm), which is the size of most repeated yarn patterns. Our Young's modulus is in \(N/(mm^2)\), and our energy is in \(N*mm\) or \(10^3*Joule\).
%\subsection{Matching Physical Yarns}
%We use the yarn pattern geometry and yarn properties provided by \citet{sperl2022}. Assuming that the yarn cross section is always circular, we can compute the Young's modulus and yarn radius directly from the data as well as some estimate of the Poisson's ratio.
\setlength{\intextsep}{5pt}%
\setlength{\columnsep}{5pt}%
\begin{wrapfigure}{R}{0.2\textwidth}
\centering
\includegraphics[width=0.2\textwidth]{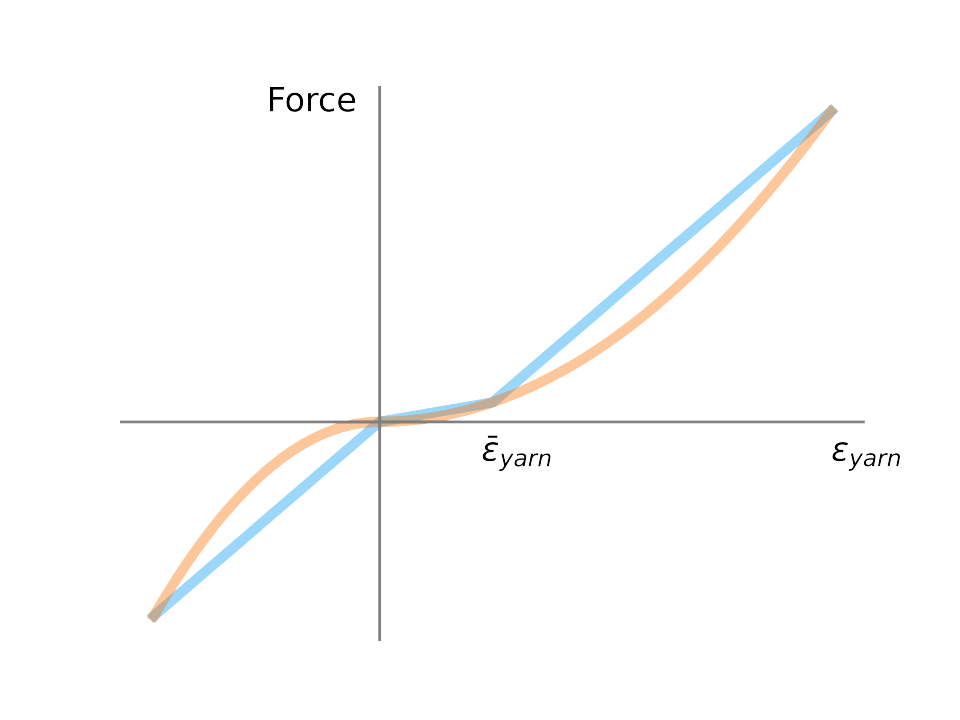}
\vspace{-0.5cm}
\caption{   \label{fig:young_modulus} Strain-force plot. Blue: Biphasic piecewise linear stretch proposed by \citet{sperl2020hylc}. Orange: Ours.}
\end{wrapfigure}
%\el{Define $\epsilon_s$ here, and ideally change either this or the one in Eq. (7)}
\begin{equation}
    E=\begin{cases}
  k_1 \epsilon_{yarn} & \text{for } \epsilon_{yarn}<0\\    
  k_2 \epsilon_{yarn} & \text{for } \epsilon_{yarn}>0   
\end{cases} \Rightarrow \nabla U_{\text{yarn}}\propto \begin{cases}
    k_1 \epsilon_{yarn}^2\\% & \text{for } \epsilon_s<0\\    
  k_2 \epsilon_{yarn}^2% & \text{for } \epsilon_s>0
\end{cases}% = {\for }
\end{equation}%\todo{Illustration: Piecewise linear (3 pieces) vs quadratic curves, one half in positive strain and the other in negative strain (inverted)}

The rate of change in the Young's modulus \(k_1,k_2\) are derived from tensile testing on various materials including polyester, yarn and wool fibers \cite{serra2019modeling}. Note that we only provide a rough estimate for the ranges of the Young's moduli, which are distinct for the materials in our scope. More accurate estimates can be obtained by conducting physical experiments on the individual yarns composing the respective fabrics.

%\todo{reference to polyester,yarn,wool material properties}. and Poisson's ratio of the fiber materials from \todo{reference: something like https://www.servicethread.com/blog/most-common-industrial-yarn-sizes}. In addition, we account for the stretch scaling effect of %Note that the Young's modulus is a rough estimate and the exact properties can be obtained from stiffness measurements as in \citet{sperl2022} given individual strands of yarns that are not fabricated, although the yarn properties might be different upon being fabricated.
%We also obtain the woven or knitted pattern of yarns given close-up captures from the front and back of the fabric and use the corresponding yarn geometry from the database provided by \citet{leaf2018interactive}. For more complex yarn structures such as a 3D woven reliefs, CT scanning is required to reconstruct the yarn geometry.

%In addition, we require an estimate of the yarn size or radius \(r\), which can be derived from the yarn denier which is defined as the mass per 9,000 meters and can be found in garment labels.
% Currently we only support fabrics made of the same material. 

% \begin{equation}
% \begin{aligned}
% \label{eqn:yarn_density_from_denier}
% r &= \sqrt{\frac{\text{denier}}{9000\pi\rho_{yarn}}}
% %K(\text{stiffness}) &= A(mm^2)\ E(N/mm^2)\\
% %E &= \frac{K}{A}\\
% %&= \frac{K}{\pi r^2}
% \end{aligned}
% \end{equation}

We initialize the yarn centerlines using the yarn pattern data from \citet{leaf2018interactive}, which contains the geometry of a single periodic repeat. Finally, the yarn radius is estimated as 
{\small $ r = \sqrt{\frac{\rho_{\text{shell}} p_x p_y}{\rho_{\text{yarn}} L_0\pi}}$}. 
%\begin{equation}
%\label{eqn:yarn_density_from_shell_density}
%    \begin{aligned}
%        r&=\sqrt{\frac{\rho_{\text{shell}} p_x p_y}{\rho_{\text{yarn}} L_0\pi}}.
%    \end{aligned}
%\end{equation}
Here, \(p_x, p_y\) is the width and length of a single repeat and \(L_0\) is the total rest length of yarns, all of which are calculated from the initialized yarn geometry. \(\rho_{\text{shell}}\) is measured by weighing a piece of fabric and dividing the mass by the number of repeats in the swatch. %we estimate the mass of a single repeat by weighing the entire fabric and dividing by the number of repeats. Subsequently, we compute

%Our parameter estimation approach has several assumptions. All of our measured fabrics are 100\% made of the same material and composed of 2D yarn structures that can be easily identified through close captures. We do not work with mixed materials or complex 3D yarn structures.

%In addition, we use weave patterns from \citet{leaf2018interactive}, which only provides the yarn geometry and yarn radius. We estimate the Young's modulus and Poisson's ratio for different types of yarns.

%|   | Young's Modulus (MPa) | Poisson's Ratio |
%|-------|-------|-------|
%| Polyester | 1531 | 0.25 |
%| Cotton | 4980 | 0.36 |

\subsection{Periodic Yarns}
%We look up the yarn geometry from a collection of common knitted and woven structures provided by \cite{leaf2018interactive}. %We treat a single 2D yarn patch as a repeating unit by imposing boundary conditions. Effectively modeling it as if it were part of a much larger patch. We do this by accounting for the yarn connections in a periodic tile by setting up the corresponding constraints. This enables us to efficiently simulate very large tiles while only accounting for the DOFs pertaining to a single repeating patch.
We model 2D fabrics as a tiling of a small repeated unit which are strongly connected when tiled periodically. Our homogenization process gathers deformation responses from a large tile by simulating the periodically repeating unit, similar to \citet{sperl2020hylc}. We model the connections of the yarn patch to adjacent \emph{ghost} tiles using a constraint formulation. We find all pairs of connected yarn ends which are in the form of {\small \((\bm{p}_i^j,\bm{p}_k^l,x_\textnormal{off},y_\textnormal{off})\)} where {\small \(\bm{p}_i^j\) } and {\small \(\bm{p}_k^l\)} are the connected yarn points (e.g. \(j\)th sampled point of the \(i\)th yarn), and \((x_\textnormal{off}, y_\textnormal{off})\) indicate the position of the ghost tile where \(v_k^l\) resides relative to the simulated tile, i.e. {\small \((x_\textnormal{off},y_\textnormal{off}) \in\{(1,0),(-1,0),(0,1),(0,-1)\}\)}. %In order to identify the yarn connections, we first compute the size of the repeat \((p_x, p_y)\).
% The size is calculated by taking the difference between the maximum and minumum coordinates of all yarn sampled points.
%We then loop through all the yarn ends and identify all pairs of endpoints that are \(p_x\) or \(p_y\) apart. If there are dangling yarns, we reduce \(p_x\) and \(p_y\) and re-run until all yarn ends are connected.
%the form of
% \begin{equation}
% \begin{aligned}
% \label{eqn:yarn_connection}
% &(\bm{p}_i^j,\bm{p}_k^l,x_\textnormal{off},y_\textnormal{off})\ \ \text{where}\\
% &(x_\textnormal{off},y_\textnormal{off}) \in\{(1,0),(-1,0),(0,1),(0,-1)\}
% \end{aligned}
% \end{equation}
Each yarn connection introduces two ghost edges \(\bm{e}_i^j\) and \(\bm{e}_k^{l-1}\) adjacent to 
{\small \(\bm{p}_i^j\) } and {\small \(\bm{p}_k^l\) } respectively
% \(\bm{e}_i^j=(\bm{p}_i^j,\bm{p}_k^{l+1}+\bm{v}_\textnormal{diff})\) and 
% \(\bm{e}_k^{l-1}=(\bm{p}_k^l,\bm{p}_i^{j-1}-\bm{v}_\textnormal{diff})\) where
% \(\bm{v}_\textnormal{diff} = p_x x_\textnormal{off}\bar e_x + p_y x_\textnormal{off} \bar e_y\). 
, corresponding to stretching-shearing constraints {\small \(\bm{C}_s\left(\bm{p}_i^j, \tilde{\bm{p}}_k^{l+1}, \bm{q}(\bm{e}_i^j)\right)\)} and {\small \(\bm{C}_s\left(\bm{p}_l^k,\tilde{\bm{p}}_i^{j-1}, \bm{q}(\bm{e}_k^{l-1})\right)\)} as well as the bending-twisting constraint {\small \(\bm{C}_b\left(\bm{q}(\bm{e}_i^{j-1}), \bm{q}(\bm{e}_k^{l})\right)\)} following Equations~\ref{eqn:stretch_shear}~and~\ref{eqn:bend_twist}. Note that $\tilde{\bm{p}}$ indicate positions in the adjacent ghost tiles. 
% \begin{equation}
% \begin{aligned}
% &\bm{C}_s\left((\bm{p}_i^j, (\bm{p}_k^{l+1}+\bm{v})_\textnormal{diff}, \bm{q}(\bm{e}_i^j)\right)\\
% &\bm{C}_s\left((\bm{p}_l^k,\bm{p}_i^{j-1}-\bm{v}_\textnormal{diff}, \bm{q}(\bm{e}_k^{l-1})\right)\\
% &\bm{C}_b\left((\bm{q}(\bm{e}_i^{j-1}), \bm{q}(\bm{e}_k^{l})\right)
% \end{aligned}
% \end{equation}
% \begin{equation}
% \begin{aligned}
% &\bm{C}_s\left((\bm{p}_i^j, \tilde{\bm{p}}_k^{l+1}+\bm{v}, \bm{q}(\bm{e}_i^j)\right)\\
% &\bm{C}_s\left((\bm{p}_l^k,\tilde{\bm{p}}_i^{j-1}, \bm{q}(\bm{e}_k^{l-1})\right)\\
% &\bm{C}_b\left((\bm{q}(\bm{e}_i^{j-1}), \bm{q}(\bm{e}_k^{l})\right)
% \end{aligned}
% \end{equation}
%\joy{Too verbose - move this section to supplementary }
%\joy{Probably don't need to write the math}

\subsection{Yarn Contacts}
At each simulation step, yarn self-collisions and inter-object collisions are processed by detecting and culling potential collisions using bounding volume hierarchies that comprise of all edge segments of the discretized yarns and the colliding triangle meshes. Edge-edge and point-triangle contacts are identified by continuous and proximity collision detection. The contacts are then resolved through a unilateral distance constraint with position-based friction to maintain a minimal separation distance equal to the yarn diameter \cite{miles2014realtime}.

\subsection{Surface Tiling}
\label{sec:surface_tiling}
%\joy{Explain extension to the RVE constraint to get rid of the effect of surface resolution}
%\begin{itemize}
%\item Periodic boundary condition: identify yarn connections
%\item RVE constraint, zero twist constraint
%\end{itemize}
We introduce additional homogenization constraints such that the periodic yarn tiling matches a deformed surface during simulation.% The tiling process is illustrated by the blue block in Figure~\ref{fig:surface-tiling}. %Note that we only simulate the periodic patch at the center of the grid an that the surface gradient is constant across the entire shell.% covering the entire shell. %We introduce two sets of constraints

\noindent \textbf{Mid-Surface Constraints.} We build on Eq.~\ref{eqn:rve_constraint} and transform all the yarn-level fluctuations to a common frame defined by the mid-surface \cite{sperl2020hylc}. The transformation matrix is essentially the rotation \(\bm{R}\) of the surface normal from the undeformed surface to the deformed surface. %Note that the periodic yarn structure only spans a sub-region of the entire shell. For shells under large out-of-plane bending deformations, the surface tangents would be different from one discretized region to another, therefore we need to compute the rotation matrices for all the copies of the periodic yarn patch covering the entire surface.
For an arbitrary yarn \(i\) with \((N + 1)\) sampled points, we compute the rotation matrix corresponding to every yarn sampled point by taking the polar decomposition of the local mid-surface deformation \(\nabla_j\boldsymbol{\phi}_i=\bm{R}_{ij} \bm{S}_{ij}\) where \(\bm{S}_{ij}\) represents the in-plane deformation. The homogenization constraint is
\vspace{-0.2cm}
\begin{equation}
\label{eqn:corotated_rve}
\begin{aligned}
\bm{C}_{\text{RVE}}(\tilde{\bm{u}})&=\sum_{j=0}^N {\bm{R}_{ij}}^T \tilde{\bm{u}}_{ij}=0%\ \text{where}\\
%\nabla_j\boldsymbol{\phi}_i&=\bm{R}_{ij} \bm{S}_{ij}
\end{aligned}
\end{equation}
\vspace{-0.2cm}

We also make sure that the boundaries are well-aligned when being tiled. For each yarn connection involving points \((\bm{p}_i^j,\bm{p}_k^l)\),
%\joy{They might have different rotation matrices! Move the \(R_{new}\), \(R_{old}\) notation here. Also emphasize how the bend-twist constraint changes to reflect this.}
%\vspace{-0.5cm}
\begin{equation}
\label{eqn:corotated_rve_boundary}
\begin{aligned}
%Egor: I replaced _b with _\Gamma to avoid confusion with "bending" constraints
\bm{C}_{\text{RVE}\_\Gamma}(\tilde{\bm{u}})={\bm{R}_{ij}}^T \tilde{\bm{u}}_{ij}-{\bm{R}_{kl}}^T \tilde{\bm{u}}_{kl}=0
\end{aligned}
\end{equation}
%\vspace{-0.5cm}
Here, \(\bm{R}_{ij}\) remains the same across the tiling, given our assumption of constant surface gradient.

\noindent \textbf{Zero-Twist Constraints.} We enforce each yarn to have zero net twist and the yarn ends to have matching twist angles, to ensure the twists do not get accumulated when tiling the yarns. The twist angle is proportional to the last entry of the Darboux vector in Eq.~\ref{eqn:bend_twist}. For a yarn with \(N\) edges:

% * The net twist of each yarn should be zero.
\vspace{-0.2cm}
\begin{equation}
\begin{aligned}
C_{\text{zt}} &= \sum_{i=0}^{N-2} \frac{\Im_2(\bar{\bm{q}}_i \bm{q}_{i+1})}{\Re(\bar{\bm{q}}_i \bm{q}_{i+1})}=0, \\
C_{\text{zt}\_\Gamma} &=\frac{\Im_2(\bar{\bm{q}}_0 \bm{q}_1)}{\Re(\bar{\bm{q}}_0 \bm{q}_1)}-\frac{\Im_2(\bar{\bm{q}}_{N-2} \bm{q}_{N-1})}{\Re(\bar{\bm{q}}_{N-2} \bm{q}_{N-1})}=0.
\end{aligned}
\end{equation}

\subsection{Yarn-Level Simulation}
\label{sec:yarnsim}
%\joy{Two-phase simulation: Write algorithm! Explain hard constraints (RVE, pin) and soft constraints.}

We execute yarn-level simulation in two phases. The first phase brings the yarns to a relaxed state. The second then minimizes the yarn energy density subject to the homogenization constraints. 

\noindent \textbf{Yarn Relaxation.} We minimize the energy in Eq.~\ref{eqn:yarn_energy} with slight modifications to the yarn constraints as shown in Eq.~\ref{eqn:phase1_rod_constraints}. We introduce a small amount of yarn shrinkage by setting the rest lengths of all yarn segments to be shorter than their initial lengths, so that the relaxed yarns are touching while the overall size of the patch remains unchanged. In practice, we set a shrinking factor \(r_s\) of 0.8 for knitted fabrics and 0.9 for woven fabrics. In addition, we assume that the yarns are naturally straight.
\begin{equation}
\vspace{-0.15cm}
\begin{aligned}
\label{eqn:phase1_rod_constraints}
\bm{C}_{s}'(\bm{p}_{i},\bm{p}_{i+1},\bm{q}_i)&=\frac{\bm{p}_{i+1}-\bm{p}_{i}}{r_sl_0}-\bm{d}_3(\bm{q}_i)\ \text{where}\ r_s\in(0,1)\\
\bm{C}_{b}'(\bm{q}_{i},\bm{q}_{i+1})&=\frac{\Im\left(\bar{\bm{q}}_i\bm{q}_{i+1}\right)}{\Re\left(\bar{\bm{q}}_i\bm{q}_{i+1}\right)}.
\end{aligned}
\end{equation}

%\gene{Should we use the following to match Eq.~\ref{eqn:darboux}? The rest Darboux vector can be ignored as we assume the yarns are naturally straight.}
%\joy{yes thank you!}

\noindent \textbf{Homogenization.} We use the relaxed yarns from the first phase as the new rest state. From this, we compute the new rest lengths and curvatures. Subsequently, we tile the yarns on a deformed surface following Sec.~\ref{sec:surface_tiling} sampled from a range of deformations detailed in Sec.~\ref{sec:homogenization}. In contrast to the compliant yarn constraints, the homogenization constraints are modeled as hard constraints in our simulations. We minimize the energy (Eq.~\ref{eqn:yarn_energy}) and record the resulting energy density:% \(\overline{U}=\frac{U}{A}\) where \({A}=p_x p_y\) is the area covered by a single periodic repeat.
\begin{equation}
\begin{aligned}
\overline{U}_{\text{yarn}} &= \frac{1}{A}\min_{\tilde{\mathbf u}}U_{\text{yarn}}\ \text{where } A=p_x p_y\\ \ s.t.\ \bm{C}_{\text{contact}}&=\bm{C}_{\text{RVE}}=\bm{C}_{\text{RVE}\_\Gamma}=\bm{C}_{\text{zt}}=\bm{C}_{\text{zt}\_\Gamma}=\bm{0}.
\end{aligned}
\end{equation}
%We refer to \citet{miles2016xpbd} for the simulation procedure with Gauss-Seidel updates. 
Although we employ a compliant constraint formulation in our implementation, we only obtain the minimized energy from the final quasi-static state. Any alternative quasi-static solver would suffice.

%\subsubsection{Yarn Shrinkage}

%Most yarn models assume that yarns are inextensible, i.e. the stretching stiffness would be very big. In practice, however, we want the relaxed yarns to be touching without changing the size of the patch, as illustrated in the RVE above. A common practice is to shrink the yarns during simulation, by setting a rest length that is slightly shorter than the initial length. The shrinking factor is usually 0.8 for knitted fabrics and 0.9 for woven fabrics.

 %Egor renamed this to "fitting" from "homogenization". reserving the word "homogenization" to building the connection between fundamental forms and yarn deformations.
\subsection{Fitting}
\label{sec:homogenization}
% We wish to homogenize the yarn and shell level models following \citet{sperl2020hylc}. Specifically, we implement a range of deformations following the first and second fundamental forms, simulate yarns subject to the deformations, record the energy density upon convergence, and optimize for the stiffness coefficients in the shell model to match the yarn-level energy density. We use the same deformation for the yarn and shell simulations, that is, \(\bm{I}=\overline{\bm{I}}\) and 
% \(\bm{II}=\overline{\bm{II}}\).

% The deformations on the yarn level are implemented as follows:

% \begin{enumerate}
%     \item Create an undeformed surface grid
%     \item Tile yarns on the undeformed surface, record the coordinates of yarn sampled points, compute the yarn rest length
%     \item Deform surface under some \((\bm{I},\bm{II})\)
%     \item Update yarn positions by looking up the coordinates on the deformed surface grid
% \end{enumerate}

We collect the yarn responses from two distinct sets of experiments where we apply either in-plane or out-of-plane deformations. % following the parametrization by \citet{sperl2020hylc}. 
%The in-plane deformations are represented by a combination of weft stretching, shearing and warp stretching that correspond to the unique entries of the symmetric matrix \(\I\).
For membrane deformations, we sample pure weft/warp stretching (\(\I_{12}=0\), \(\I_{11}\neq 1\) and/or \(\I_{22}\neq 1\)) and pure shearing (\(\I_{11}=\I_{22}= 1\), \(\I_{12}\neq 0\)) to avoid ambiguity in the off-diagonal entries of the transformation matrix. For bending deformations, in addition to sampling orthogonal sets of singly-bent surfaces where only one of \(\mathbb{II}_{11}\) and \(\mathbb{II}_{22}\) is non-zero, we also sample bending along the bias direction with \(|\II_{12}|=|\II_{11}|=|\II_{22}|\neq 0\). %Our sampled strains are in the range \([0,0.5]\) for stretching and \([-0.5,0.5]\) for shearing and bending, which accounts for the amount of elongation of our yarn materials before breaking.
% \begin{itemize}
%     \item In-plane stretching: \((s_x,0,s_y)\) where \(s_x,s_y\in [1, 2.25]\), corresponding to strain values in \([0, 0.5]\)
%     \item In-plane shearing: \((1,\mathbb{I}_{12},1)\) where \(\bm{I}_{12}\in [-0.5,0.5]\)
%     \item Bending along weft: \((\bm{II}_{11},0,0)\) where \(\bm{II}_{11}\in [-1,1]\)
%     \item Bending along warp: \((0,0,\bm{II}_{22})\) where \(\bm{II}_{22}\in [-1,1]\)
%     \item Bending along bias: \((\bm{II}_{12},\bm{II}_{12},\bm{II}_{12})\) and \((-\bm{II}_{12},\bm{II}_{12},-\bm{II}_{12})\) where \(\bm{II}_{12}\in [-1,1]\)
% \end{itemize}
% \begin{table}[]
%     \centering
%     \begin{tabular}{ccc}
%        Pure Stretch & Pure Shear & \\
%        $\begin{bmatrix}
%            \sqrt{\mathbb{I}_{11}} &  & \\  & \sqrt{\mathbb{I}_{22}} & \\ & & 1
%        \end{bmatrix}$ & $\begin{bmatrix}
%            1 & \mathbb{I}_{12} & \\  & \sin{(\cos^{-1}(\mathbb{I}_{12})}) & \\ & & 1
%        \end{bmatrix}$ & \\
%        Weft Bend & Warp Bend & Bias Bend \\
%     \end{tabular}
%     \caption{Caption}
%     \label{tab:my_label}
% \end{table}
Given the minimized energy densities corresponding to all sampled surface deformations, we use the weighted least squares method to optimize the compliance matrices for stretching/shearing and bending for the shell model. Our weights follow a Gaussian distribution centered at the zero strain to produce a better fit at small strains as these are more common due to the limited extensibility of yarns. Given \(N\) samples,
%\el{Here $C_s$ and $C_b$ are used but not the same as in eq (16). Need to reformulate (8) to make it easier to write this}
%fit a linear/nonlinear orthotropic StVK model.
\begin{equation}
\label{eqn:least_squares}
\begin{aligned}
\epsilon = \min_{\bm{\gamma}}&\sum_{i=0}^{N-1} w_i(\overline{U}_{\text{shell}}^i(\bm{\gamma}) - \overline{U}_{\text{yarn}}^i)^2\ \text{where}\\
\overline{U}_{\text{shell}}^i(\bm{\gamma}) &= \frac{1}{A}(U_{\text{stretch}}^i + U_{\text{bend}}^i)\\
&=\frac{1}{2A}\overline{\bm{C}}_s^T\overline{\bm\alpha}^{-1}_s\overline{\bm{C}}_s+\frac{1}{2A}\overline{\bm{C}}_b^T\overline{\bm{\alpha}}^{-1}_b\overline{\bm{C}}_b,\\
\end{aligned}
\end{equation}
$\bm{\gamma} = (\coeff_{00}, \coeff_{11}, \coeff_{22}, \coeff_{01}, \bcoeff_{00}, \bcoeff_{11}, \bcoeff_{22})$ is the set of stiffness coefficients for stretch and bending, and
\(\overline{\bm{\alpha}}^{-1}_s,\overline{\bm{\alpha}}^{-1}_b,\overline{\bm{C}}_s,\overline{\bm{C}}_b\) are defined in Section~\ref{sec:shell_stvk}. We do not jointly sample membrane and bending deformations, i.e. \(\I=\bm{0}\) if \(\II\neq\bm{0}\) and vice versa.

\section{Results}  \label{sec:results}
\begin{figure}[t]
    \centering
    \includegraphics[width=0.99\linewidth]{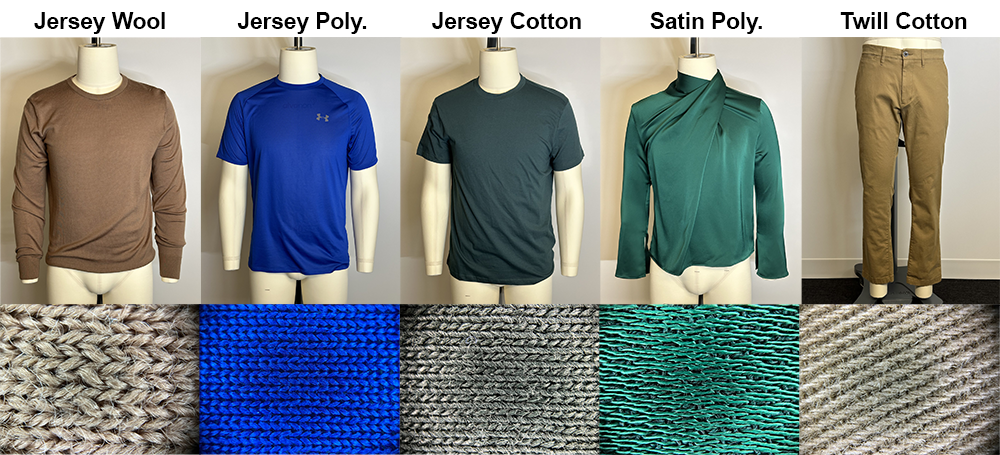}
    \caption{The garments used in our results. The bottom row shows close-up images of the fabrics.
    \label{fig:materials}}
    \vspace{-0.2cm}
\end{figure}

We estimate the shell parameters for five real fabrics comprised of different yarn structures and materials, as shown in Figure \ref{fig:materials}. We derive the yarn parameters for three distinct materials: wool, cotton and polyester, following Section \ref{sec:yarn_param_estimation}, and provide the exact parameters and simulation setup in a supplementary document. %Noting that each yarn material has a maximum amount of stretch \(\epsilon_{max}\) it can withstand before breaking. We compute the rate of change in Young's modulus as {\small \(k_1=\frac{E_{max}-E_{min}}{\epsilon_{max}}\)}, and do not consider yarn shrinkage in the homogenization process since the yarns are initially relaxed in all our simulations.
The virtual cloth is visualized using a surface-based shading model~\cite{zhu2023realistic} with the required normal and tangent information being directly inferred from the relaxed woven and knitted patterns.
%\begin{table}[]
%    %\centering
%    \setlength\tabcolsep{0.01pt}
%    \begin{tabular}{ccccc}
%    Jersey Wool&Jersey Poly.&Jersey Cotton&Satin Poly.&Twill Cotton\\
%     \hspace{-0.25cm}\includegraphics[width=0.1\textwidth]{img/fabrics/jersey-wool.jpg}\hspace{-0.25cm}\vspace{-0.2cm} &
%     \includegraphics[width=0.1\textwidth]{img/fabrics/jersey-poly.jpg}\hspace{-0.25cm}&
%    \hspace{-0.25cm}\includegraphics[width=0.1\textwidth]{img/fabrics/jersey-cotton.jpg}\hspace{-0.25cm} & 
%    \includegraphics[width=0.1\textwidth]{img/fabrics/green-satin.jpg}\hspace{-0.2cm}&
%    \includegraphics[width=0.1\textwidth]{img/fabrics/twill-pants.jpg}\\ 
%    \hspace{-0.25cm}\includegraphics[width=0.1\textwidth,height=0.08\textwidth]{img/fabrics/jersey-wool-closeup-reduced.jpg}\hspace{-0.25cm} &
%     \vspace{-0.2cm}\includegraphics[width=0.1\textwidth,height=0.08\textwidth]{img/fabrics/jersey-poly-closeup-reduced.jpg}\hspace{-0.25cm}&
%    \hspace{-0.25cm}\includegraphics[width=0.1\textwidth,height=0.08\textwidth]{img/fabrics/jersey-cotton-closeup-reduced.jpg}\hspace{-0.25cm} & 
%    \includegraphics[width=0.1\textwidth,height=0.08\textwidth]{img/fabrics/green-satin-closeup-reduced.jpg}\hspace{-0.2cm}&
%    \includegraphics[width=0.1\textwidth,height=0.08\textwidth]{img/fabrics/twill-pants-closeup-reduced.jpg}
%    \end{tabular}
%    %\vspace{-1.5cm}
%        \begin{figure}[H]
%    \caption{The fabrics used in our results.}
%        \label{fig:materials}
%    \end{figure}
    %\vspace{0.5cm}
    %\renewcommand{\tablename}{Fig.}
%\end{table}

% \bgroup
% \def\arraystretch{1.2}
% Moved the following to supplementary material
% \begin{table}
%     \begin{tabular}{ccccc}
%     \toprule
%     Material & \(\rho_{yarn}\) \((g/cm^3)\) & \(E_{min}\) (GPa) & \(E_{max}\) (GPa)& \(\epsilon_{max}\)\\
%     \midrule
%     Wool & 1.31 & 2.3 & 3.4 & 0.3\\
%     Cotton & 1.52 & 5.3 & 6.2 & 0.1\\
%     Polyester & 1.38 & 1.5 & 10.6 & 0.4\\
%     \bottomrule
%     \end{tabular}
%     \caption{Material parameters for yarn-level simulations. }
%     \label{table:fabric_properties}
% \end{table}
% \egroup
\begin{figure}[t]
    \centering
    \includegraphics[width=0.8\linewidth]{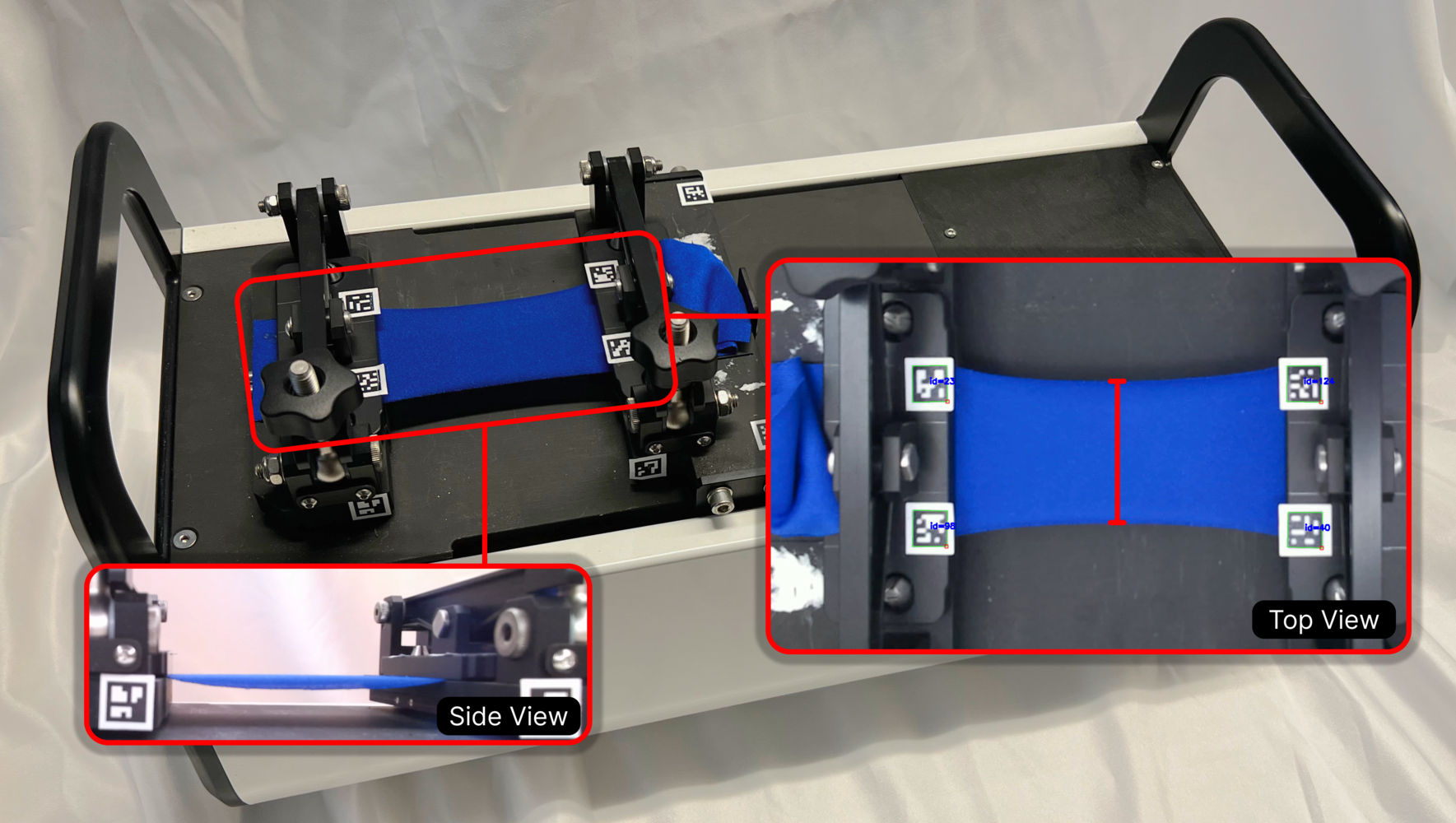}
    \caption{Fabric analyzer (FAB) measures force exerted by the fabric as it is stretched. The machine is enhanced with ArUco markers \cite{opencv} to determine the ratio between stretch (horizontal length) and orthogonal compression (vertical length drawn in red).\label{fig:fab_sample}}
\end{figure}
% \begin{figure}[t]
% \centering
%     \label{fig:stretch_test_demo}
%     \includegraphics[width=0.32\linewidth]{img/stretch_test/stretch-test-full-yarn-sim-placeholder.png}
%     \includegraphics[width=0.32\linewidth]{img/stretch_test/stretch-test-nonlinear-stvk-placeholder.PNG}
%     \includegraphics[width=0.32\linewidth]{example-image-a}
%     \caption{Stretch test on three configurations. \todo{} \label{fig:stretch_test_demo}}%Fabric analyzer (FAB) measures force exerted by the fabric as it is stretched. The machine is tracked using ArUco markers %\cite{opencv} to determine the ratio between stretch (horizontal length) and orthogonal compression (vertical length drawn in red).\label{fig:fab_sample}}
% \end{figure}

\begin{figure}
    \centering
    \includegraphics[width=\linewidth]{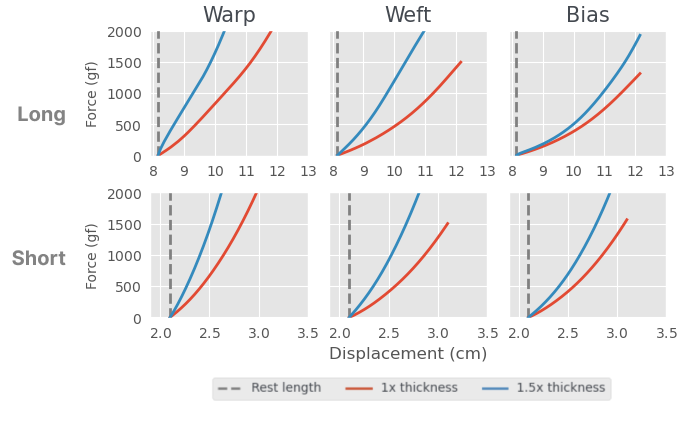}
    \caption{The force plot of the stretch test for the estimated cotton jersey material (red) and the same material with 1.5$\times$ thicker yarns (blue). As expected, the thicker material exhibits higher stress during stretching.}
    \label{fig:compare_thickness}
\end{figure}

\begin{figure*}[t]
    \centering
    \includegraphics[width=0.78\linewidth]{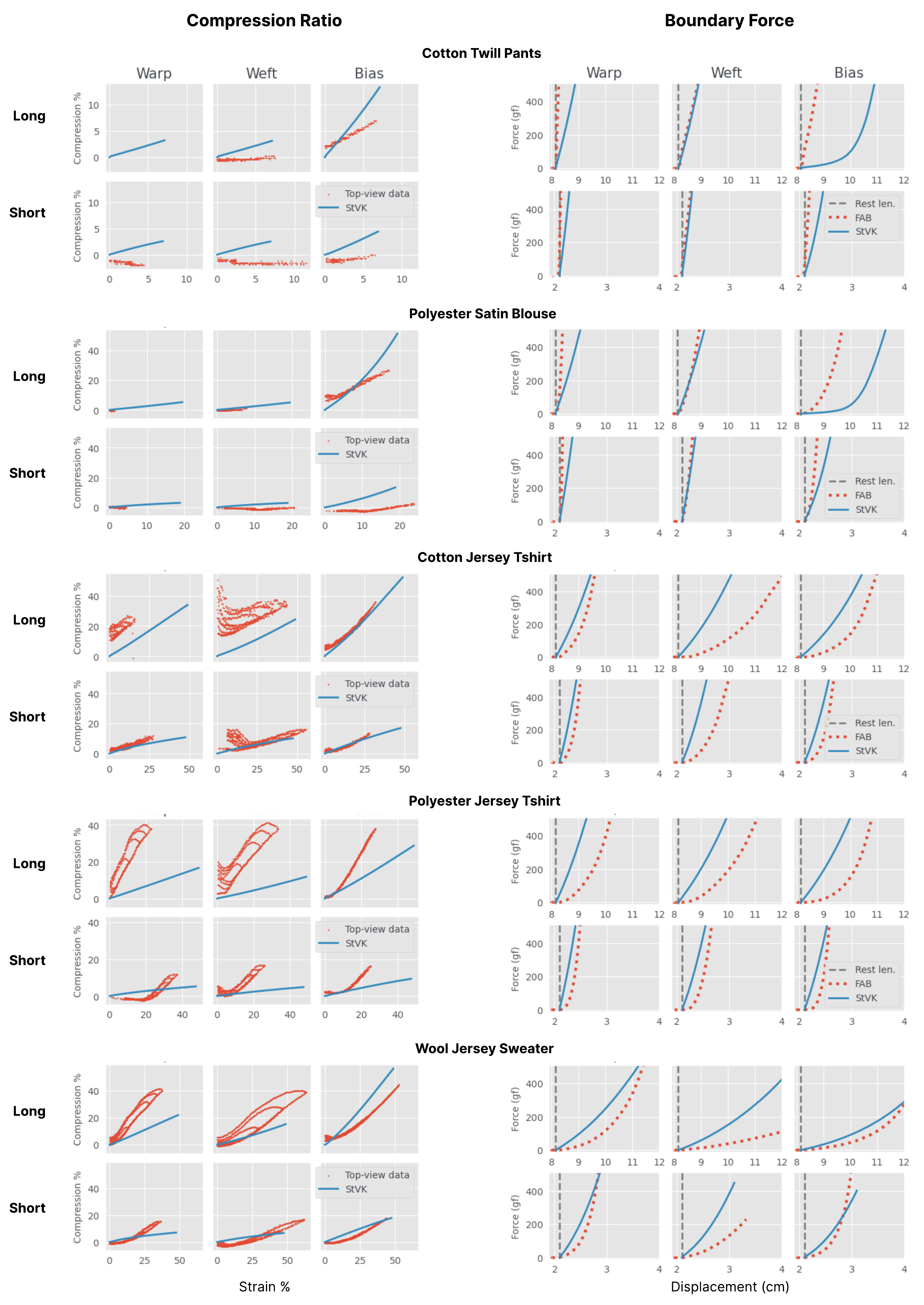}
    \caption{Compression ratio (left column) and boundary force (right column) plots for 5 different materials comparing captured data from the FAB (red dots) against shell simulation results (solid blue curves). It is remarkable to see that simulated shell materials capture the key characteristics of yarn structures such as decreased stiffness/increase compression ratio in the bias direction for woven patterns. In some cases like the cotton jersey t-shirt, the compression ratio (determined by the slope of the curve) matched the data remarkably well, in spite of there being no inverse process to match fabric measurements in our pipeline. \label{fig:plots}}
\end{figure*}

%\subsection{Validation}
% \subsection{Stretch Test}\label{sec:stretch_test}
\textbf{Stretch Test.}
We validate our results against real world fabric stretch tests, where a piece of rectangular fabric is clamped at two ends, stretched and the resulting forces on the clamps measured. We use the commercially available fabric analyzer (FAB) \cite{fab} to perform stretch tests with $8\times5$ cm and $2\times5$ cm pieces of fabric cut along the warp, weft and bias (at 45 degrees) directions. We augment the FAB with additional markers to track the strain and capture the amount of visual deformation (compression) along the orthogonal direction as pictured in \figref{fig:fab_sample}. We then reproduce the same scenario in simulation at the sheet level and plot the resulting boundary forces and compression ratios against real world measurements from the FAB as shown in \figref{fig:plots}. This test is critical in confirming that our pipeline can predict real fabric behaviors.
Furthermore, we use the same stretch test to demonstrate in \figref{fig:compare_thickness} how thicker yarns produce sheets with larger stresses, a phenomenon previously unexplored in homogenized cloth.

%\todo{Four stretch test images: Warp/Weft, full yarn sim vs. shell sim}\todo{Discussion: Green jersey cotton fits very well even without inverse pipeline}

\begin{figure*}[p]
    \centering
    %\begin{subfigure}{0.55\textwidth}
    %\centering
    %\includegraphics[width=\linewidth]{img/sphere_drape/SphereDrapes.jpg}%
    %\caption{Sphere drape example.\label{fig:sphere_drape}}%
    %\end{subfigure}%
    % \begin{subfigure}{\textwidth}
    % \centering
    \includegraphics[width=\linewidth]{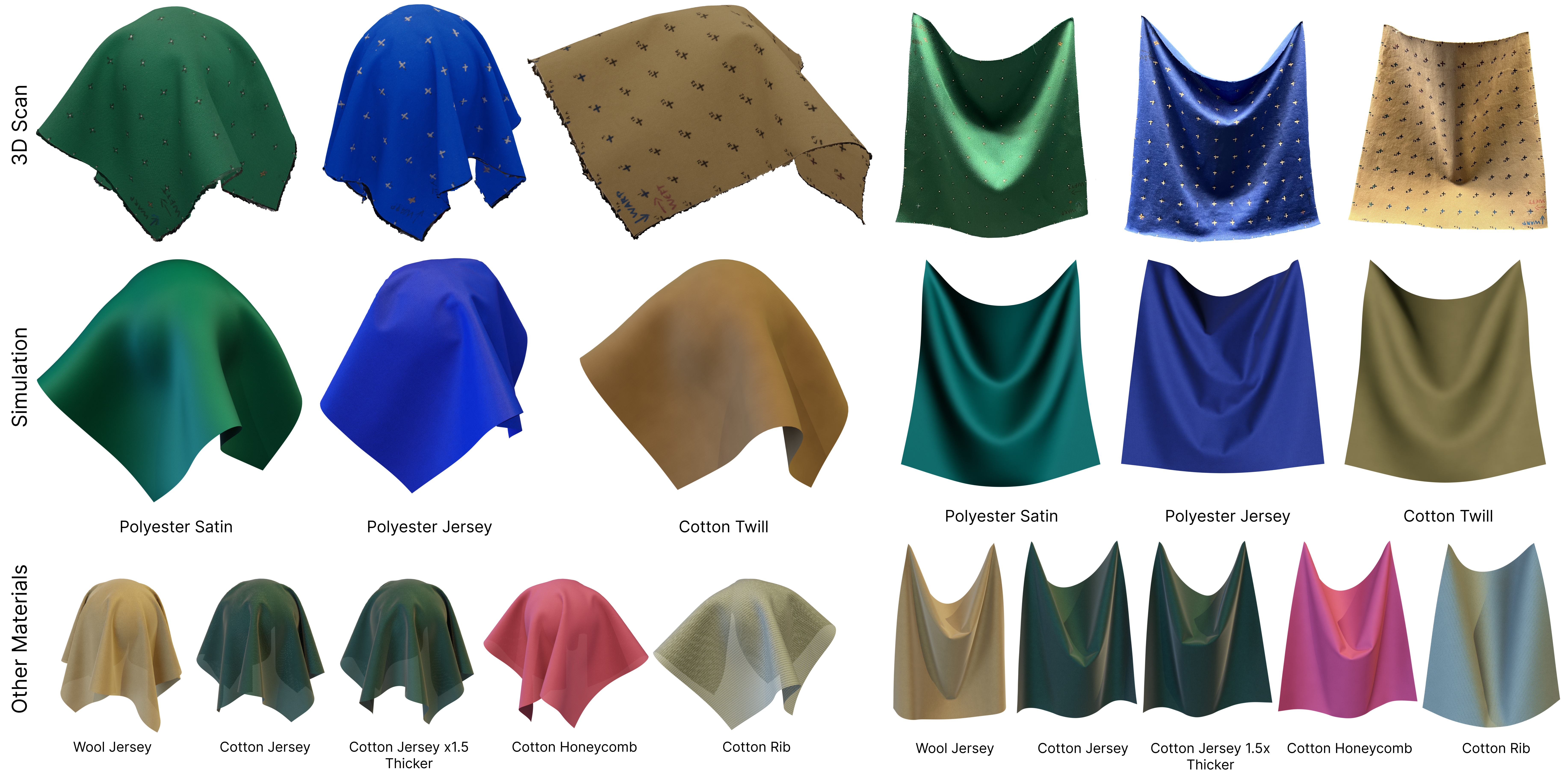}
    \caption{We show two different drape setups where we compare simulated results to a real reference, one where the swatch is draped on a sphere and a second one where the top corners are fixed. Given the use of very limited information about the fabrics, we do not expect our drapes to match the reference exactly. However, our estimated materials produce qualitatively similar results to the references such as highly anisotropic bending in satin and twill fabrics, and are noticeably distinct for different material types. \label{fig:drape}}
    % \end{subfigure}
    \includegraphics[width=\linewidth,trim=0 0 0 -10]{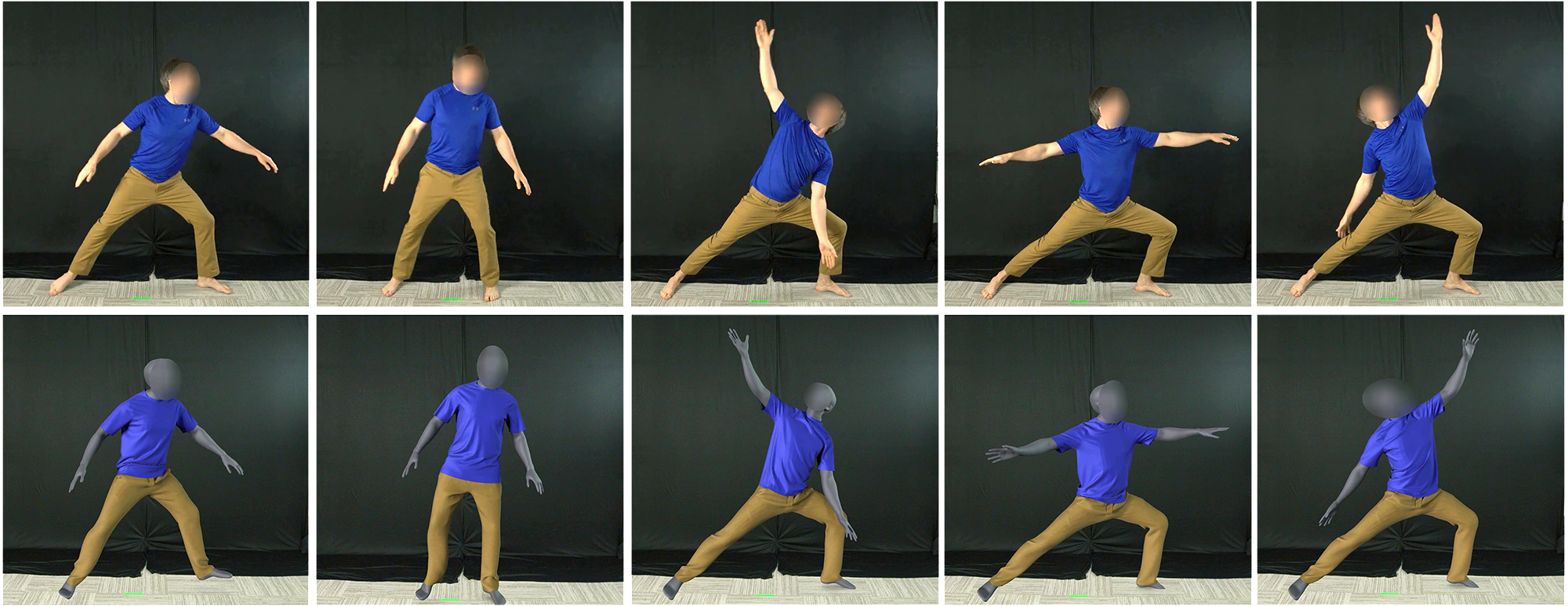}
    \caption{Full body cloth simulation of the \emph{cotton twill pants} and \emph{polyester jersey t-shirt} with parameters estimated using our pipeline. The simulated pants exhibit dramatic creases in poses 3, 4 and 5 matching the reference. The simulated T-shirt exerts fairly natural wrinkles in response to the arm movements and body twists, which are most evident in poses 3 and 5.\label{fig:full_body}}
\end{figure*}

%We conduct a three-way comparison among stretch tests performed on (i) full yarn representations (without periodic boundary conditions) of the entire fabric, (ii) homogenized StVK shell models and (iii) real fabrics. The results are shown in \figref{fig:stretch_test_demo}.
% \subsection{Sphere Drape}
\textbf{Drape Tests.}
The stretching quality of materials can be verified numerically by comparing force measurement as in \figref{fig:plots} (right column). However, for bending, the measured forces are very small and require very sensitive force sensors. While the FAB produces some force output for bending tests, it is unreliable for some materials as reported by Browzwear's software. To validate bending parameters, we drape a piece of cloth on a sphere, reconstruct geometries using a standard photogrammetry software (e.g. Agisoft Metashape) and compare the result to the same scenario reproduced in simulation as shown in \figref{fig:drape}. We also perform draping against a wall and compare to captures of real drapes, demonstrated in the same figure.

% \subsection{Full Body Validation}
\textbf{Full Body Validation.}
To demonstrate how our models are realized in real world applications, we simulate clothing on a full body.
An assembled virtual t-shirt and pants are draped on an animated body and compared to a reference video of the same outfit as shown in \figref{fig:full_body}. Our simulated garments exhibit realistic folds and creases in response to the body movements.

% \begin{itemize}
% \item Yarn simulation: Rod model (Romero's cantilever and bend-twist tests)
% \item Full pipeline: Stretch test and comparison between 1. full yarn pattern; 2. captured data (camera); 3. homogenized sheet
% \begin{itemize}
%     \item Visual: Stretch/compression ratio
%     \item Force-deformation plots
% \end{itemize}
% \item Demo: Drape shell on squared table, and we should observe bias bending
% \item Full body simulation results
% \end{itemize}

\section{Limitations and Future Work} \label{sec:limitations}
%\begin{wrapfigure}{R}{0.25\linewidth}
%    \centering
%    \includegraphics[width=\linewidth]{img/curling_cropped.png}
%    \caption{Curling effect underneath the fabric hidden from the top-view camera.\label{fig:curling}}
%\end{wrapfigure}

\textbf{Edge curling effects.} While we use top-view captures for estimating compression ratio, they are insufficient for estimating in-plane compression data due to curling effects at the boundaries of the stretched fabric. The side view close up in \figref{fig:fab_sample} shows fabric at the edges bending out of the view from the top-view camera. To remedy this, we propose to add heterogeneous curved rest shape near the boundary and measure curling amount from a side-view capture. We believe this discrepancy is in part responsible for the mismatch in compression ratio shown in the left column in \figref{fig:plots} for jersey fabrics.

\begin{wrapfigure}{R}{0.25\linewidth}
    \centering
    \includegraphics[width=\linewidth]{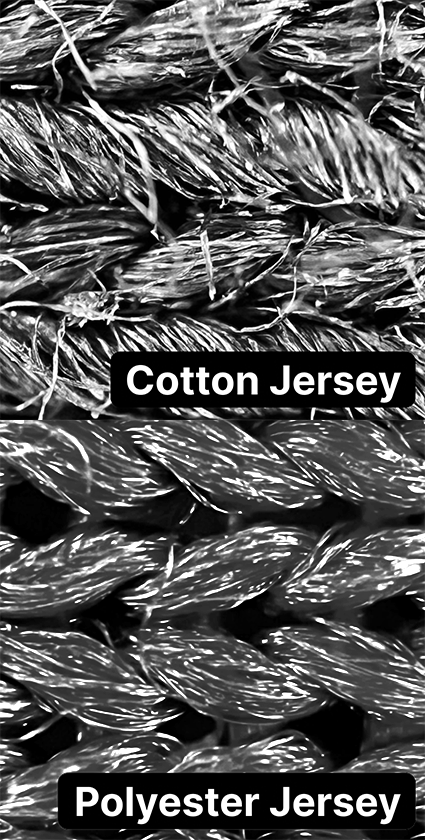}
    % \caption{Closeup photograph of cotton and polyester materials showing the difference in yarn texture.}
    % \label{fig:closeup}
\end{wrapfigure}
\textbf{Internal friction.} We apply a uniform friction coefficient for all materials in our yarn simulation, while the internal frictions of different materials are vastly divergent in real fabrics. Moreover, we do not estimate frictions between yarn fibres, which causes hysteresis effects in deformed fabrics. Consistent with previous work,
we discover a considerable amount of yarn behavior caused by friction, especially in the knitted fabrics. Internal frictions also cause unwanted yarn sliding and lead to mismatches between simulated results and measured data, which explain the difference between compression ratios on cotton and polyester knits, since polyester yarns have fewer dangling fibers as shown in the inset. %,\joy{What does this mean? Can we say more locally stable?} and tend to exhibit lower inter-yarn friction. due to yarn shiftingregardless of hysteresis since friction will change the maximal amount that yarns will shift. This

%\textbf{Damping.}

\textbf{Fabric materials.} Our yarn-level parameter estimation assumes that the yarns are homogeneous, i.e. the entire fabric is 100\% composed of the same yarn material. In practice, however, a blend of organic and synthetic materials is more commonly used in clothing such as pants to improve the durability while preserving comfort. Our experiment uses twill pants made of 98\% cotton and 2\% spandex, which exhibit higher contraction and flexural rigidity compared to pure cotton \cite{almetwally2014effects}, while our estimated shell parameters assume 100\% cotton. Furthermore, the fabrication process alters the physical properties of the fabric. For example, manufacturing pants requires a finishing process of pressing and ironing, which changes the stretching/shrinkage properties of the underlying fabric at high temperatures \cite{islam2019consequences}. Both factors contribute to the significant quantitative mismatch between simulated and real twill pants in the stretch test. Future work entails distinguishing between different yarn materials in a composite via CT scanning and accounting for the post-processing stage in the yarn simulation. %Since our parameter estimation relies on individual yarn properties and limited information on the fabric, the . %Composite yarns often exhibit vastly different physical properties compared to which results in the most amount of mismatch across all materials in our experiments.

\section{Conclusion} \label{sec:conclusion}
We propose an end-to-end method for estimating shell model properties corresponding to real fabrics. Through simple measurements of the fabric swatches and close-up captures, we effectively obtain information about the yarn structures and physical properties such as yarn thickness. Combined with textile research literature that provide approximations for the tensile strengths of different materials, we initialize yarn-level models with properties derived from the real fabrics, which are preserved through our yarn-shell homogenization pipeline. Subsequently, we sample a range of surface deformations given by the first and second fundamental forms, record the responses of periodic yarn patterns on each sampled deformed surface in terms of energy density, and optimize all the stiffness parameters of an orthotropic StVK shell model, including off-diagonal bending terms representing warp-weft coupling. We validate our result quantitatively and visually against measurements and captures of real fabrics on stretch tests and drape experiments, and demonstrate a good quantitative approximation and qualitative match with real fabrics such as reproducing the anisotropic behaviors in satin and twill weaves, despite the lack of an inverse process. Our homogenized shells not only establish the key characteristics of each yarn pattern, but also showcase the effects of yarn material properties. Using only physics-based simulation, our method provides reasonably accurate initial estimates for parameter estimation pipelines requiring more comprehensive knowledge of real fabrics.

%%
%% The acknowledgments section is defined using the "acks" environment
%% (and NOT an unnumbered section). This ensures the proper
%% identification of the section in the article metadata, and the
%% consistent spelling of the heading.
% \begin{acks}

% \end{acks}

%%
%% The next two lines define the bibliography style to be used, and
%% the bibliography file.
\bibliographystyle{ACM-Reference-Format}
\bibliography{ref}
\clearpage

%%
%% If your work has an appendix, this is the place to put it.
% \appendix

\end{document}